**UNIVERSIDAD CENTRAL DE LAS VILLAS**

**FACULTAD DE MATEMÁTICA, FÍSICA Y COMPUTACIÓN**

**DEPARTAMENTO DE FÍSICA**

# LA FOTO-BIOFILIA DE LA VÍA LÁCTEA

Tesis presentada en opción al grado científico de Doctor en Ciencias Físicas.

## OSMEL MARTÍN GONZÁLEZ

**Santa Clara**

**2010**

UNIVERSIDAD CENTRAL DE LAS VILLAS

FACULTAD DE MATEMÁTICA, FÍSICA Y COMPUTACIÓN

DEPARTAMENTO DE FÍSICA

# LA FOTO-BIOFILIA DE LA VÍA LÁCTEA

Tesis presentada en opción al grado científico de Doctor en Ciencias Físicas.

Autor: Lic. OSMEL MARTÍN GONZÁLEZ

Tutores: Dr. Rolando Cárdenas Ortiz

Dr. Jorge Ernesto Horvath

**Santa Clara**

**2010**

DEDICATORIA

… , a mis dos amorosas princesitas: Lianet y Liset

## SINTESIS

En el presente trabajo se aborda de manera integral las implicaciones que tiene el régimen fotobiológico y particularmente la radiación ultravioleta (RUV) en el origen, extensión y evolución de la vida fotosintética en nuestra galaxia. En esta concepción se abordan elementos a escalas espacio temporales notablemente diferentes que incluyen desde aspectos cosmológicos del problema, pasando por aspectos de la estabilidad estelar y planetaria, hasta considerar elementos de la biosfera a una escala mucho más regional o local pudiendo alcanzar, incluso, el nivel celular. Un énfasis particular se presta al papel que pueden potencialmente desempeñar las explosiones estelares y particularmente los brotes de rayos gamma BRG sobre la biosfera terrestre. En este punto, se discuten algunos criterios de daño por radiación UV establecidos en la literatura y se introducen algunos criterios adicionales. Aspectos característicos de estas perturbaciones a nivel de ecosistema son discutidos cuantitativamente para el caso de algunos ecosistemas acuáticos, resaltando las complejidades inherentes a los mismos. Para finalizar, se introducen algunas nociones sobre la extensión e importancia de los ritmos circadianos para los seres vivos permiten un acercamiento a aquellas teorías que abordan los problemas relacionados con el propio proceso del surgimiento de la vida y su definición.

INDICE

# Introducción

# INTRODUCCIÓN

El surgimiento de la Vida ha sido, sin lugar a dudas, uno de los tópicos más polémicos y menos comprendidos en toda la historia de la humanidad. Con raíces tan antiguas como la propia civilización humana y alcanzando representaciones que van desde lo meramente artístico o religioso hasta alcanzar aspectos fundamentales de carácter filosófico, el tema aparece como un elemento recurrente en casi todas las culturas.

Sin embargo, la inclusión formal de este tópico en el campo de la ciencia no sería probablemente hasta el año 1862 con los trabajos del químico francés Louis Pasteur. En ellos y bajo condiciones de laboratorio bien controladas, Pasteur refutaría definitivamente las ideas sobre la generación espontánea demostrando la utilidad del método científico y particularmente del enfoque experimental en este tipo de estudios. Aunque los trabajo de Pasteur constituyeron un notable paso de avance sus resultados confirmaron el carácter sumamente complejo de la Vida y la imposibilidad, al menos desde el punto de vista científico, de ofrecer una respuesta simple para explicar su origen.

En correspondencia con las dificultades anteriores, los estudios sobre el origen de la Vida en la Tierra se dividen en lo que podríamos llamar dos corrientes fundamentales; las que consideran a la Vida como un fenómeno originario de nuestro planeta y las que no.

Dentro del marco de las que consideran a la vida como un fenómeno originario se encuentran las teorías de la evolución química como son el "Mundo ARN" y "El metabolismo primero" (consultar Lazcano, 2010). En este contexto, la vida surge a partir de elementos químicos sencillos, que van generando bloques moleculares cada vez más complejos. Así la materia evolucionaría desde las formas inorgánicas inertes hasta la materia orgánica viva.

Dentro de las teorías que consideran la vida como un proceso no originario se encuentra la Panspermia, introducida originalmente por el también químico y ganador del premio Nobel Svante Arrhenius en 1903. Su dogma central es que la vida se formó inicialmente en el espacio exterior pudiendo llegar a nuestro planeta gracias a los meteoritos u otros objetos que colisionaron con la Tierra sobre los eones Hadeico o Arcaico. Estos objetos portarían células u organismos simples, que fueron depositados sobre la superficie de la Tierra, y que se revitalizarían cuando las condiciones ambientales fueran las adecuadas, desarrollándose, sembrando la vida en el planeta y evolucionado a lo largo de millones de años para dar lugar a las formas actuales de vida. Aunque esta teoría no resuelve directamente el problema del surgimiento de la vida, el hallazgo de moléculas orgánicas complejas en el espacio y la composición química de meteoritos como el de Murchison han hecho que esta teoría se reconsidere una y otra vez.

Por otro lado, tanto en correspondencia con las teorías de la evolución química como en el de la panspermia, las condiciones climáticas de la llamada tierra temprana (eones Hadeico o Arcaico) desempeñan un papel fundamental para comprender los procesos de instauración, diseminación y posterior evolución de la vida en nuestro planeta. Numerosos son los trabajos referidos en la literatura especializada sobre esta temática (Grenfell y colaboradores, 2010; Kasting y Catling, 2003). Dentro las variables climáticas consideradas, una buena parte de estos trabajos se centran en las condiciones radiativas de la atmosfera arcaica, el hábitat y los posibles niveles de irradiación (particularmente en la región del RUV) a que estaban sometidos estos primeros organismos (Cockell, 2000; Cockell y Raven, 2007).

El interés motivado por la RUV se debe en buena medida a los efectos directos que pueden ocasionar la exposición a niveles moderados de RUV sobre los organismos vivos y fundamentalmente sobre el material genético. Dicha influencia ha motivado que algunos autores la consideren como uno de los posibles promotores de la evolución biológica (ejemplo ver Hessen, 2008). Sin embargo, es de esperar que dicho papel haya variado sensiblemente durante toda la historia biogeoquímica de nuestro planeta. Probablemente los altos niveles de la RUV en el eón arcaico hayan ejercido como un factor más bien restrictivo a la vida durante esta etapa, condición esta que pudo suavizarse ya en el proterozoico con niveles de $O_2$ alrededor del uno por ciento que garantizaban un bloqueo parcial por parte del ozono. Por otro lado, si bien el incremento en los niveles de oxígeno en la atmósfera disminuye sensiblemente los niveles de RUV, este incremento favorece el daño foto-oxidativo por formación de radicales libres. Este parece un hecho notable si tenemos en cuenta que la llamada explosión del Cámbrico (~550 Ma), periodo en el cuál se diversificó considerablemente el número de especies que habitaban el planeta, coincide aproximadamente con la instauración de la atmósfera moderna.

De las consideraciones anteriores se reafirma el criterio de considerar la vida como un fenómeno complejo, no solo en su esencia, sino por la continua co-evolución con otros factores que conforman el clima y le confieren la mayor parte de las propiedades distintivas que exhibe nuestro planeta. Este hecho, fundamenta una buena parte de los programas que se llevan a cabo hoy en día con el objetivo de detectar vida en otros planetas como es el caso de las misiones "Terrestrial Planet Finder" y Darwin ( Kaltenegger, 2010). En este sentido, considerar la Tierra como un exo-planeta permite definir y calibrar criterios específicos para medir o inferir algún nivel de actividad biológica. Entre los criterios más aceptados se encuentran la presencia de determinados gases como el oxígeno, metano y óxido nitroso, así como la presencia del llamado límite rojo alrededor de 0.7 um, típico de la extensión de las formas de vida fotosintéticas (Consultar por ejemplo DesMarais y colaboradores, 2002).

Por otro lado, pese al papel preponderante del sol sobre nuestro planeta, no es descartable considerar el efecto estocástico de otras fuentes de radiación a escala galáctica. Un cúmulo importante de investigaciones en los últimos años han sido dirigidas en este sentido. El objetivo fundamental de estos trabajos consiste en estimar los posibles impactos sobre la atmósfera y biosfera de nuestro planeta de eventos astrofísicos muy intensos, así como sus posibles consecuencias sobre el proceso de evolución biológica (Smith y colaboradores, 2004b). Dentro de este grupo se destacan por su intensidad las explosiones de supernovas (ESN) y los llamados brotes de rayos gamma (BRG) (Thomas y colaboradores, 2005; Galante y Horvath, 2007). Por la magnitud de estos eventos y teniendo en cuenta el carácter mutagénico o incluso letal que pueden tener los altos niveles de RUV existen hipótesis que los relacionan con algunas de las grandes extinciones masivas que han tenido lugar en el pasado de nuestro planeta (Mellot y colaboradores, 2004). La importancia de estos eventos puede trascender el propio marco de la Tierra y ser un elemento negativo a considerar en la habitabilidad de determinada región como en el caso de la zona galáctica habitable introducida por (González y colaboradores, 2001) y explorada por otros autores como (Lineweaver y colaboradores, 2004)

**La novedad y actualidad del tema**

Este trabajo se enmarca en una serie de investigaciones y estudios recientes centrados fundamentalmente en la influencia potencial que pueden ejercer tanto la radiación solar como la asociada a eventos astrofísicos intensos como son las explosiones estelares sobre el medio ambiente y la biosfera. En él se abordan varios elementos y herramientas que permiten una mejor comprensión y evaluación de los efectos nocivos asociados al incremento de la radiación ultravioleta (RUV) y permiten abordar problemas científicos de gran vigencia como el manejo, cuidado y protección de ecosistemas o el cambio climático global. En este contexto se discuten además aspectos referentes a la complejidad intrínseca de los sistemas biológicos y el carácter fenomenológico del propio proceso de modelación.

Por otro lado, el trabajo a diferencia con otros, ayuda a establecer de manera integral el papel de la radiación sobre el origen, distribución y evolución de la vida alcanzando incluso una escala cosmológica. Se proponen escenarios probables para la llamada tierra temprana como el eón Hadeico o el principio del Arcaico, aspectos discutidos por su importancia en la literatura especializada. En el trabajo además, se analizan aspectos relacionados con la propia esencia de la vida como fenómeno auto-organizado y lejos del equilibrio. En este sentido, se discute la importancia de los ritmos circadianos y de las oscilaciones bioquímicas de manera general con la introducción de un modelo dinámico simple para la célula primitiva.

Los resultados fundamentales que conforman esta tesis se encuentran publicados en:

1) Martín O., Cárdenas R., Guimarais M., Peñate L., R. Horvath J.E., Galante D. ´´Effects of gamma rays bursts in Earth´s biosphere''. Astrophysics and Space Science (2010) 326: 61–67

2) <u>Martín O.</u>, Peñate, L., Alvaré, A., Cárdenas R., Horvath J.E., Galante D. ''Some possible constraints for life's origin''. Origins of Life and Evolution of Biospheres (2009) 39 (6): 533-544

3) <u>Martín, O.</u>, Galante, D., Cárdenas, R. Horvath, J.E ''Short-term effects of gamma rays bursts on Earth''. Astrophysics and Space Science (2009) 321: 161–167

4) Sussman R., Quiros I., <u>Martin O.</u> ''Inhomogeneous models of interacting dark matter and dark energy''. General Relativity and Gravitation 37 (12): 2117-2143 (2005)

Adicionalmente sobre esta temática se encuentra aún en proceso de arbitraje el trabajo (por invitación) para conformar el capítulo de un libro sobre la génesis de la Vida que publicará la editora Springer

5) <u>Martín O.</u>, Peñate, L., Cárdenas R., Horvath J.E. The Fotobiological regime in the very early Earth and the emergence of life. Para capítulo del libro ''Origins of Life'' de Springer.

**El objeto de la investigación:** El entorno de radiación en nuestra galaxia y su influencia sobre las formas de vida fotosintéticas y sobre la biosfera de manera general.

**Sus objetivos:**

- Estimar las condiciones del Universo en general, y de la Vía Láctea en particular, que contribuyen a la aparición de la vida.

- Estimar las repercusiones fundamentales que tiene el régimen fotobiológico y sus posibles alteraciones sobre los productores primarios fotosintéticos.

- Explorar el comportamiento de las biosferas planetarias ante grandes perturbaciones del régimen fotobiológico asociados a la ocurrencia de eventos estelares intensos como las explosiones estelares.

**La hipótesis de trabajo:** El régimen fotobiológico y sus alteraciones constituyen un elemento clave para comprender el origen, evolución y distribución de la vida, no solo en el ámbito terrestre, sino en un contexto general. Por tanto el Universo en general, y en particular nuestra galaxia, muestran un determinado grado de foto-biofília.

**El fundamento metodológico y los métodos utilizados para realizar el trabajo de investigación**: Dado lo interdisciplinar del tema, se hizo una extensa revisión bibliográfica y se consultó a diversos especialistas nacionales y extranjeros, para posteriormente seleccionar las herramientas que se usaron para cumplir los objetivos. Estas herramientas estaban dentro de diversas áreas, tales como Astrofísica Teórica, Física Atmosférica (interacción radiación-atmósfera), Ecología Teórica (dinámica de ecosistemas bajo estrés radioactivo), Fotobiología (estimación de daño biológica de las radiaciones UV) y Ciencia de la Complejidad (probables efectos no lineales en el comportamiento de la biosfera, tanto a nivel de ecosistema como celular).

# LA VIDA FOTOSINTÉTICA EN EL CONTEXTO COSMOLÓGICO

# 1 LA VIDA FOTOSINTÉTICA EN EL CONTEXTO COSMOLÓGICO

El objetivo de este capítulo es examinar las potencialidades del Universo en general, y de nuestra galaxia en particular, para la emergencia de vida fotosintética.

## 1.1 Modelo general de Vida

Astrobiología es el nombre moderno con que se designa la ciencia que estudia el origen, evolución, distribución y destino de la vida en la Tierra y en el Universo. Es un área interdisciplinar donde confluyen Astronomía, Física, Biología, Geología y otras ciencias.

Para estudiar la vida en el contexto cosmológico, es necesario guiarse por un modelo de vida lo más general posible, que rebase las limitaciones de la vida tal y como la conocemos en la Tierra. Nos adherimos a un modelo en que como mínimo se exigen tres condiciones para la emergencia de vida:

a) Elementos químicos biogénicos. Ejemplo: C, H, O, N

b) Solvente en que los elementos puedan mezclarse y reaccionar para formar moléculas biológicas complejas. Ejemplo: $H_2O$

c) Una fuente de energía que guíe la bioquímica mencionada en el inciso anterior. Ejemplo: Luz y el proceso de fotosíntesis en los productores primarios en la Tierra

Es bueno señalar que en la Tierra, si bien todas las especies conocidas se incluyen en los ejemplos de los dos primeros puntos, no todas las formas de vida dependen de la luz solar y la fotosíntesis (los seres humanos dependemos de la fotosíntesis indirectamente, mediante la ingestión de alimentos). Los organismos quimiosintéticos que viven en los respiraderos hidrotérmicos en las profundidades oceánicas no dependen en lo absoluto de la luz solar, ya que ellos elaboran sus alimentos a partir de sustancias químicas presentes en tales ecosistemas. Así las cosas, en nuestro planeta conocemos dos formas de vida:

a) Vida que depende de la luz (directa o indirectamente): Organismos fotosintéticos (fitoplancton, algas, plantas superiores) y animales herbívoros, carnívoros, detritívoros, omnívoros, etc.

b) Vida independiente de la luz: Organismos quimiosintéticos (los cuales representan la minoría en nuestro planeta).

En este trabajo nos concentramos en el estudio de la vida dependiente de la luz, o sea, dependiente del proceso de fotosíntesis. Esta elección responde fundamentalmente a la notable repercusión de este proceso en la evolución biogeoquímica de nuestro planeta y a la disponibilidad de energía. Actualmente la fotosíntesis supera en varios órdenes a otras formas de producción primarias y se estima que en la tierra temprana, alrededor de 380 Ma, su contribución a la productividad global fuera similar a la de los organismos quimiosintéticos (Raven 2009; Canfield, 2006).

1.2 **Una perspectiva cosmológica al problema**

Al menos en principio, es posible abordar el problema del surgimiento de la vida y particularmente de la vida fotosintética desde una perspectiva tan general como la cosmológica (Davies, 2004). Si uno examina cuidadosamente los requerimientos asumidos en la sección anterior, salvo la condición de existir un solvente adecuado, las naturaleza de las restantes dos condiciones pueden trascender el marco puramente estelar o galáctico donde se desarrolla la vida. Desde esta perspectiva, dichas condiciones aparecen ligadas directamente a las leyes más generales que rigen nuestro universo y que han permitido la formación de estructuras tales como las propias galaxias, estrellas, planetas, nubes moleculares gigantes alrededor de las cuales se originara y evolucionara la vida.

En este sentido, una importancia particular se le confiere a los procesos de formación de estrellas debido a sus múltiples roles en el contexto astrobiológico. Además de ser la fuente de energía por excelencia para el proceso de fotosíntesis y proveer una plataforma climática relativamente estable, en su interior se sintetizan los elementos químicos indispensables (biogénicos) que son expulsados violentamente en las etapas finales de su ciclo de vida. De hecho, las abundancias actuales de estos elementos en el Universo, en la galaxia y particularmente en el sistema solar implican la existencia de otras generaciones de estrellas. Se estima que estas primeras estructuras se formaron a expensas del material primigenio (H y He fundamentalmente) y se caracterizaron por ser extremadamente masivas y de gran luminosidad así como por tener ciclos de vida extremadamente cortos del orden de unos 3 millones de años. El estudio y caracterización de estas primeras formaciones constituye hoy en día un campo activo de investigación tanto desde el punto de vista observacional como teórico (Consultar por ejemplo Yoshida y colaboradores, 2003; Wise y Abel, 2008).

Es conveniente señalar que el proceso de formación de elementos químicos también ocurrió en los primeros segundos de nuestro universo en la llamada nucleosíntesis primordial. Sin embargo, las condiciones de este proceso favorecieron mayoritariamente la formación de los elementos químicos más ligeros del sistema periódico tales como el hidrógeno, el helio y en menor cuantía el litio. Elementos más pesados como el carbono o el nitrógeno quedaron limitados únicamente al nivel trazas haciendo despreciable su contribución frente a los procesos ordinarios de nucleosíntesis estelar.

**1.2.1 La expansión del Universo y el proceso de formación de estructuras**

En el marco del modelo cosmológico estándar, el universo actual no es más que el resultado de un largo proceso evolutivo (expansión adiabática) a partir de un estado inicial, caracterizado por una densidad y temperatura extremas. En correspondencia con este modelo, la forma en que se ha llevado el proceso de expansión del Universo es clave para comprender los procesos de formación de estructuras y por consiguiente para estimar la biofilia de este. Típicamente, para un universo plano, homogéneo e isótropo, la tasa de expansión se expresa mediante el llamado parámetro de Hubble $H$:

$$H = \frac{\dot{a}}{a} \qquad (1.1)$$

donde $a$ es el factor de escala del Universo. Definamos el contraste de densidad $\delta$ como

$$\delta = \frac{\delta\rho}{\langle\rho\rangle} \qquad (1.2)$$

donde $\rho$ es la densidad dentro de la sobredensidad, $\langle\rho\rangle$ es la densidad promedio del Universo y $\delta\rho = \rho - \langle\rho\rangle$. La ecuación relativista de evolución de las sobredensidades es del tipo oscilador amortiguado (suponiendo $H$ aproximadamente constante):

$$\ddot{\delta} + 2H\dot{\delta} - 4\pi G\delta = 0 \qquad (1.3)$$

donde $G$ es la constante gravitacional. El segundo término del miembro izquierdo es el análogo a la fricción viscosa y realmente contribuye a la amortiguación de las sobredensidades. Si la expansión universal tuviera una tasa por encima de cierto umbral $H_u$ (dependiente del modelo cosmológico), entonces la formación de estructura no hubiese procedido, o sea, no se hubieran formado galaxias, nubes moleculares gigantes etc. Lo que sucede en la realidad es que la magnitud $H$ es variable en el tiempo, dependiente del modelo cosmológico y del inventario total de materia y energía del universo. Para el modelo cosmológico estándar, que presupone que a gran escala el universo es homogéneo e isótropo, la ecuación (3) tiene dos posibles soluciones: una en la que sobredensidad es eliminada conforme el tiempo avanza según una relación del tipo ($\delta \approx 1/t$); en la otra la sobredensidad crece proporcional a la expansión ($\delta \approx a$) contribuyendo al proceso de formación de estructuras hasta conformar el Universo en su apariencia actual.

1.2.2 **El proceso de formación de estructuras desde una perspectiva local**

Si bien el principio cosmológico estándar ha resultado una herramienta extremadamente útil para estudiar nuestro universo, su validez se encuentra limitada a escalas considerablemente grandes del orden de 100-300 Mpc. A estas escalas, el universo se puede considerar prácticamente homogéneo y su dinámica se encuentra determinada fundamentalmente por la influencia de un fluido exótico mayoritario de presión negativa: la energía oscura. Este componente, de naturaleza aún poco comprendida, constituye prácticamente dos tercios del contenido material de universo y se considera el responsable de su expansión acelerada en correspondencia con los datos observacionales provenientes de las supernovas tipo Ia.

Por otro lado, para realizar estudios a escalas inferiores resulta imposible obviar el carácter no homogéneo en la distribución de la materia agrupada en galaxias, cúmulos y supercúmulos de galaxias. A estas escalas la dinámica aparece dominada por otro componente exótico mayoritario: la materia oscura. Ocupando prácticamente un tercio de todo el contenido material del universo la materia oscura, a diferencia de la energía oscura, parece agruparse significativamente por debajo de la llamada escala de homogeneidad (100-300 Mpc) y sus efectos pueden ser notables incluso en la dinámica galáctica. Aún sin un candidato bien establecido, la materia oscura se considera el ingrediente indispensable para explicar la dinámica de las grandes formaciones estelares así como el comportamiento anómalo encontrado en las curvas de rotación típicas de muchas galaxias incluyendo la nuestra.

Con vistas a modelar la evolución de la sobredensidad a nivel local (un observador en ella) Sussman, Quirós y el autor (Sussman, Quirós y Martín, 2005) proponen un modelo alternativo que generaliza en varios aspectos trabajos anteriores sobre esta temática. La construcción del modelo descansa sobre el caso esféricamente simétrico de las llamadas soluciones de Szafron-Szekeres para las ecuaciones de Einstein y reconoce en su definición una interacción explícita entre los dos componentes mayoritarios: la energía oscura (modelada como un fluido homogéneo) y la materia oscura (considerada como un fluido no homogéneo).

La aplicación un esquema como este presenta algunas ventajas pues permite reducir formalmente el estudio de la sobredensidad, dependiente del tiempo y de las coordenadas espaciales, a un escenario clásico del tipo FLRW de mucha menor complejidad al depender únicamente de la variable temporal. El método permite además explorar el impacto que pueden tener a nivel local varios formalismos plausibles en la literatura para la energía oscura como es el caso de los campos escalares. Para este fin basta solo con proponer una relación adecuada entre la densidad y la presión (ecuación de estado) para describir el comportamiento de la energía oscura. La elección conveniente de la dependencia espacial y las condiciones de frontera permite describir la dinámica de agrupamientos de materia oscura a una escala arbitraria que ajusten convenientemente a las restricciones observacionales.

Por otro lado, el hecho de proponer una interacción explicita (no gravitacional) entre la materia oscura y la energía oscura puede ser un elemento cuestionable dentro del modelo. Sin embargo, la naturaleza relativamente desconocida de ambos componentes podría justificar en alguna medida este tipo de suposición. De cualquier manera, un escenario como este podría ser interesante, tanto en el ámbito teórico como práctico, y apropiado para describir el embrión de grandes formaciones estelares, cuna posterior de la vida. En este sentido y aunque de manera paradójica, la vida, enmarcada como un fenómeno a nivel de las interacciones electromagnéticas en el seno la materia ordinaria, emerge bajo los designios de un Universo gobernado por dos componentes de naturaleza tan exótica y tan poco comprendida como pueden ser la energía y materia oscuras.

Con vistas proseguir en nuestro análisis y aunque hasta el momento hemos dirigido nuestra atención a aspectos generales relacionados con la vida y sus condicionamientos en la gran escala, en los próximos epígrafes acercaremos nuestro estudio a escalas considerablemente menores, próximas al escenario donde la vida, tal y como la conocemos se desarrolla.

## 1.3 Emergencia de vida fotosintética en planetas tipo terrestre que orbitan estrellas medianas

En este epígrafe particularizamos el modelo de vida que utilizaremos, a partir del modelo más general esbozado en el epígrafe 1.1. Consideraremos que la lista mínima de elementos biogénicos está integrada por CHON, el solvente es el agua y la fuente de energía es la luz proveniente de una o varias estrellas. Consideraremos que la zona foto-habitable de un sistema estelar es aquella en la que el agua puede estar en forma líquida en la superficie de un planeta con presión atmosférica similar a la terrestre. Con este modelo, se considera que las estrellas medianas (F, G y K) son las más adecuadas para que en sus sistemas planetarios emerja la vida fotosintética, pues las estrellas grandes (O y B) tienen tiempos de vida relativamente cortos, del orden de millones o decenas de millones de años, como para que pueda emerger la vida. Por otro lado, las estrellas pequeñas (M) obligan a sus planetas a acercarse demasiado para que el agua pueda estar líquida en su superficie, y entonces los planetas sufren anclaje de marea, dando siempre la misma cara a la estrella. La anterior situación puede implicar congelamiento de la atmósfera en el hemisferio oscuro y el arrastre hacia éste de la atmósfera del hemisferio iluminado, lo que pudiera terminar en un estado de congelación total de la atmósfera. Sin embargo el tema no está exento de debate, si tenemos en cuenta que algunos estudios recientes sugieren que esta situación no sucedería necesariamente en todos los casos. Para una discusión más detallada de estos tópicos consultar (Lammer, 2007).

Otro elemento importante a tener en cuenta en el caso de la vida fotosintética es la temperatura de la estrella. Teniendo en cuenta que en primera aproximación los espectros de emisión de estos sistemas se ajustan aceptablemente a un cuerpo negro, la longitud de onda y por consiguiente la energía de los fotones en la región de mayor emisión va a depender directamente de la temperatura en correspondencia con la conocida ley de Wien $\lambda = \frac{b}{T}$. Esto implica, teniendo en cuenta que la reducción del $CO_2$ y el $H_2O$ en la fotosíntesis es un proceso energéticamente costoso, la necesidad de emplear etapas (reacciones fotoquímicas) adicionales con vistas a un mejor aprovechamiento de la energía. Esta consideración podría ser importante para aquellos sistemas estelares con temperaturas más bajas que nuestro Sol como es el caso de las ya mencionadas estrellas M (Wolstencroft and Raven, 2002; Cockell and Raven 2004).

En este trabajo y de manera simplificada, podríamos decir que la vida fotosintética puede emerger en planetas tipo terrestre orbitando estrellas medianas (F, G o K) y en los cuales hay agua líquida e irradiancia suficiente en la superficie planetaria como para que se pueda realizar la fotosíntesis.

1.3.1 **Emergencia de vida en la Tierra**

La historia natural de nuestro planeta se divide en eones, según se muestra en la Tabla 1.1

| Eón | Tiempo antes del presente en Ma |
|---|---|
| Fanerozoico | 542-presente |
| Proterozoico | 2500-542 |
| Arcaico | 3800-2500 |
| Hadeico | 4500-3800 |

Tabla 1.1 Historia geológica de la Tierra

El origen de la vida fotosintética en nuestro planeta es aún controversial. La mayoría lo sitúan a inicios del Arcaico (hace 3800 Ma) aunque otros prefieren una fecha menos controversial (hace 3500 Ma) (Schopf, 1987). Se ha considerado tradicionalmente qué anterior a estas fechas, durante el eón Hadeico, la Tierra era un mundo inhóspito con océanos de magma y sometido a intenso bombardeo de asteroides (Shoemaker, 1983). Sin embargo, el reciente descubrimiento de zircones (rocas que necesitan de agua líquida para formarse) que datan de ese eón ha abierto la posibilidad de que dentro del Hadeico hubiera un período de relativa calma, precisamente en la era que podríamos bautizar como Hadeico intermedio (entre 4400 y 4000 Ma antes del presente) (Valley y colaboradores, 2002). Este escenario ha sido llamado una Tierra Temprana Fría (TTF). Si realmente existió, quizá su régimen fotobiológico difería poco del que existió en el Arcaico temprano: el Sol ya había terminado su fase T de Tauro (Cnossen y colaboradores, 2007) (por lo que sus emisiones en el ultravioleta y el visible serían similares en ambas eras). Con respecto a la atmósfera, la composición exacta está aún en debate (Catling y Claire, 2005; Catling y colaboradores, 2001), aunque parece razonable suponer niveles de $N_2$ similares a los actuales asi como importantes cantidades de gases de efecto invernadero tales como $CO_2$ y $CH_4$. En ambas eras la rotación más rápida del planeta implicaría fuertes vientos superficiales de más de 150 km/h. Esta situación generaría, entre otros patrones complejos, corrientes verticales circulares (circulación de Langmuir) en el océano que se extenderían hasta decenas de metros de profundidad. Como veremos posteriormente, este hecho jugará un papel crucial en la eficiencia de la fotosíntesis.

El rango de longitudes de onda útiles para la fotosíntesis es aproximadamente el mismo que podemos ver: 400-700 nm. A esta banda típicamente se le llama Radiación Fotosintéticamente

Activa (RFA). Asumiremos que toda esta banda favorece la fotosíntesis, mientras que la radiación ultravioleta (RUV) la inhibe, debido a que causa daños en el aparato fotosintético y a que daña los ácidos nucleicos (ADN y ARN) obligando a la célula a invertir energía en repararlos. Detalles sobre estos tópicos pueden encontrarse en (Cockell, 2000; Hader y Worrest, 1991; Neale *y* colaboradores, 1993; Cullen y Neale, 1994; Vincent y Roy, 1993).

Tanto durante el Hadeico intermedio como durante el Arcaico temprano, la atmósfera carecía de bloqueadores de radiación ultravioleta como el ozono, aunque existe la hipótesis de la probable existencia de smogs de hidrocarburos (Pavlov y colaboradores, 2001). En esas condiciones es de esperar que la vida fotosintética emergiera en el agua (Cockell, 2000; Cockell y Raven, 2007), para así aminorar el crudo régimen fotobiológico existente en la superficie continental (que además solo cubría un 10% del planeta durante el Arcaico).

La parte superior del océano, en que la irradiancia es suficiente para hacer posible la fotosíntesis, es llamada zona fótica. Hoy día alcanza hasta 200 metros de profundidad en las aguas más claras en las cuencas centrales oceánicas, y hay razones para pensar que el océano Arcaico era así de transparente. La parte superior de esta zona es típicamente influenciada por la circulación de Langmuir si los vientos soplan a más de 3 m/s, algo típico en el Arcaico temprano. Estas corrientes circulares y verticales mezclan las aguas de la parte superior del océano, haciendo sus propiedades físicas (temperatura, densidad) constantes, de ahí el nombre de capa mezclada. Por debajo de ésta, la circulación es detenida por gradientes de densidad y temperatura (picnoclinas y termoclinas), y se establece una estratificación de las aguas. La profundidad de la capa mezclada hoy día es de decenas de metros, y depende de la localidad.

Considerando el crudo régimen fotobiológico durante el Hadeico y Arcaico temprano (Cockell, 2000), es razonable asumir que la biota viviente en la capa mezclada del océano fuera resistente a las radiaciones, especialmente considerando efectos como la circulación de Langmuir. Estas corrientes circulares periódicamente expondrían a los organismos unicelulares a la superficie, donde recibirían altas dosis de RUV. De modo que esperaríamos una biota de baja diversidad, debido a la restricción ultravioleta. Además, estos organismos deberían tener buenas capacidades de reparación, de modo que empleamos el llamado modelo $E$ que aparece en (Fritz y colaboradores, 2008) para estimar las tasas de fotosíntesis en la capa mezclada del océano.

La tasa de fotosíntesis $P$ en este modelo se calcula mediante:

$$P = P_{pot}\left(\frac{1}{1 + E^*_{inh}}\right) \qquad (1.4)$$

Donde $E^*_{inh}$ es la irradiancia inhibitoria adimensional, dada por la radiación ultravioleta, mientras que $P_{pot}$ es la tasa de fotosíntesis en ausencia de fotoinhibición, dada por:

$$P_{pot} = P_s \left(1 - e^{-E_{PAR}/E_s}\right) \qquad (1.5)$$

En la expresión anterior, $P_s$ es la tasa de saturación de la fotosíntesis en ausencia de inhibición; $E_{PAR}$ (en W.m$^{-2}$) es la irradiancia de la luz visible, mientras que $E_s$ (en W.m$^{-2}$) es un parámetro del modelo. La irradiancia inhibitoria adimensional de radiación ultravioleta está dada por:

$$E^*_{inh} = \sum_{\lambda=200nm}^{400nm} E(\lambda)\, \varepsilon_E(\lambda)\, \Delta\lambda \qquad (1.6)$$

Donde $\varepsilon_E(\lambda)$ (en $W^{-1}.m^2$) son los pesos biológicos que cuantifican la efectividad de la exposición espectral $E(\lambda)$ (en $W^{-2}.m^{-2}.nm$), es decir, $\varepsilon_E(\lambda)$ representa la inhibición de la fotosíntesis causada por la radiación ultravioleta de longitud de onda $\lambda$.

Sustituyendo la ecuación (1.5) en la (1.4) y normalizando respecto a $P_s$ obtenemos:

$$\frac{P}{P_S} = \frac{1 - e^{-E_{PAR}/E_S}}{1 + E_{inh}^*} \qquad (1.7)$$

Ahora queda claro que la tasa de fotosíntesis es una combinación de dos factores: el numerador en la ecuación anterior favorece la fotosíntesis (pues la irradiancia de la radiación fotosintéticamente activa está allí), mientras que el denominador la inhibe, debido a la presencia del factor inhibitorio de irradiancia ultravioleta. La emergencia de la fotosíntesis es aún un tema abierto, pero parece razonable asumir las mismas rutas fotosintéticas y aproximadamente tasas similares durante toda la historia de la vida. Esto nos permite estimar la influencia del régimen fotobiológico sobre la fotosíntesis, es decir, la influencia de diferentes irradiancias solares y tramitancias atmosféricas.

Dividimos la zona fótica en dos capas para estimar las tasas fotosintéticas: la capa mezclada y el resto de la zona fótica. La profundidad de la capa mezclada, donde existe la circulación de Langmuir, depende de la velocidad del viento que sopla en la superficie. Para el Arcaico temprano es razonable asumir profundidades iguales o mayores de 30 metros (Cockell, 2000). En este trabajo, como ejemplo para nuestros cálculos, asumimos una capa mezclada de 40 metros de profundidad, en gran medida porque en la referencia anterior hay multitud de datos hasta esa profundidad. Realmente, para calcular las tasas de fotosíntesis en la capa mezclada, en la ecuación (1.7) usamos $E_{PAR}$ y $E^*_{inh}$ como se reportan en (Cockell, 2000), y promediamos $E_s$ de los datos para fitoplancton en 16 estaciones antárticas (Fritz y colaboradores, 2008). Por otro lado, para estimar $E_{RFA}$ y $E^*_{inh}$ por debajo de la capa mezclada se extrapolaron los resultados reportados en (Cockell, 2000) hasta las profundidad de 200 m.

Cuando la irradiancia de la radiación fotosintéticamente activa $E_{RFA}$ toma el valor del parámetro $E_s$, entonces el numerador en el miembro derecho de la ecuación (1.7) se iguala a 0,63, es decir, el parámetro $E_s$ representa la irradiancia de la radiación fotosintéticamente activa $E_{RFA}$ que asegura el 63 % de la tasa máxima de fotosíntesis ($P/P_S = 63\%$). Así, mientras menor sea el valor de $E_s$, más eficiente es el organismo que realiza la fotosíntesis, puesto que alcanza el 63% de la tasa máxima con una irradiancia menor.

Podemos preguntarnos si existirían apreciables diferencias en el potencial fotosintético de los organismos del Arcaico temprano comparados con los unicelulares actuales. Para considerar esto, consideramos no sólo el valor $E_s \approx 20 W.m^{-2}$ obtenido de promediar los 16 valores reportados en (Fritz y colaboradores, 2008), sino que también usamos $E_s \approx 15 W.m^{-2}$ y $E_s \approx 25 W.m^{-2}$. El valor de $E_s$ para muchos organismos fotosintéticos actuales, incluyendo multicelulares, cae en el rango anterior. Esto puede corroborarse en el trabajo de (Biber y colaboradores, 2003) donde se reportan las curvas de fotosíntesis-irradiancia para tres grupos diferentes de macroalgas (tenga en cuenta que las unidades de irradiancia en esta referencia no son $W.m^{-2}$).

Como dijimos anteriormente, durante el Arcaico Temprano nuestro planeta completaba un giro en unas 15 horas. Una rotación tan rápida implicaría fuertes vientos superficiales y, consecuentemente, las corrientes circulares de Langmuir llegarían a decenas de metros de profundidad. Estas corrientes hundirían y subirían organismos unicelulares capturados en la capa mezclada, de manera que las irradiancias promedio recibidas por una célula durante un ciclo completo en una celda de Langmuir serían:

$$\overline{E_{inh}^*} = \frac{\sum_{z=0^-}^{z_L} \sum_{\lambda=200nm}^{400nm} E(\lambda, z)\varepsilon_E(\lambda)\Delta\lambda\Delta z}{z_L} \qquad (1.8)$$

$$\overline{E_{RFA}} = \frac{\sum_{z=0^-}^{z_L} \sum_{\lambda=400nm}^{700nm} E(\lambda, z)\Delta\lambda\Delta z}{z_L} \qquad (1.9)$$

donde $z_L$ es la máxima profundidad a la que la circulación de Langmuir se extiende. Como mencionamos anteriormente, por debajo de esta capa mezclada aparece la estratificación del agua como resultado de gradientes de temperatura y/o densidad (termoclinas y picnoclinas), lo cual no permite la circulación y mezcla.

En la Figura 1.1 presentamos curvas de la tasa de fotosíntesis relativa vs. Profundidad en el océano Arcaico, para tres valores del parámetro $E_s$. Como en (Cockell, 2000), consideramos ángulos solares zenitales de 0 y 60 grados (ver figura 1.2). El primero proporciona máxima irradiación con el Sol directamente sobre nuestras cabezas, mientras que el segundo da irradiancias similares a las actuales durante el mediodía en altas latitudes.

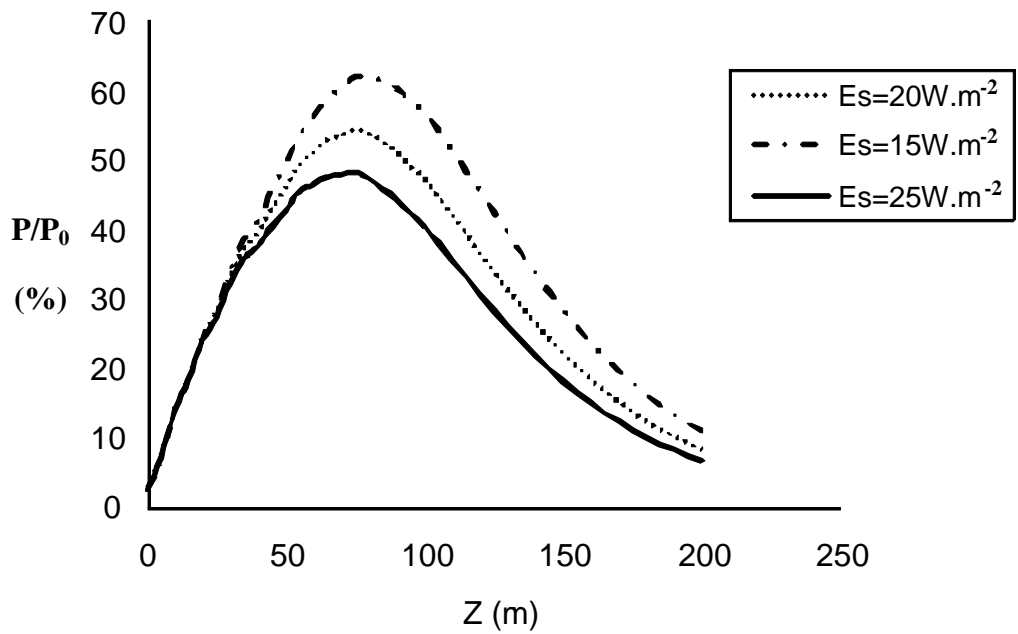

Fig. 1.1 Tasa relativa de fotosíntesis vs. profundidad en el océano Arcaico, para (azs = $0^0$)

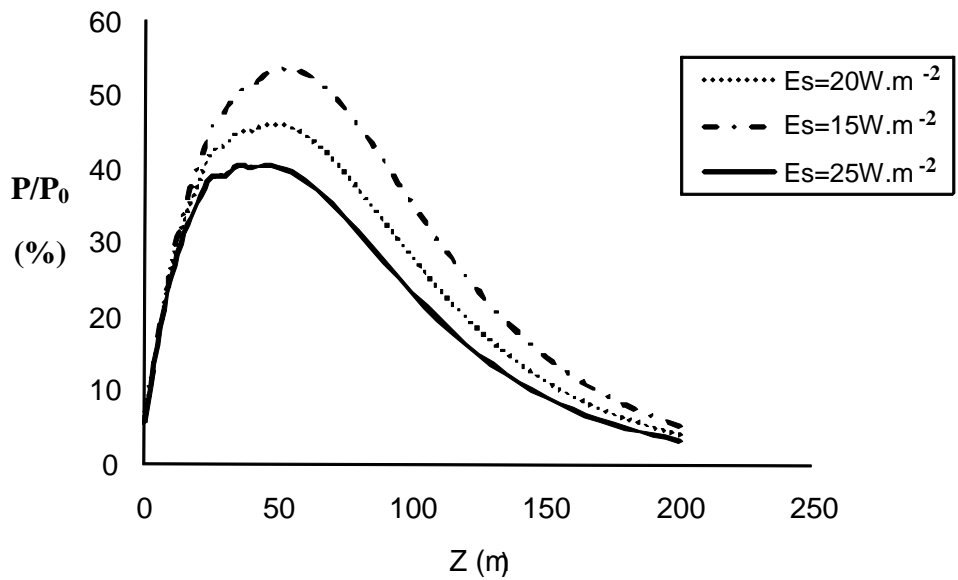

Fig. 1.2 Tasa relativa de fotosíntesis vs. profundidad en el océano Arcaico para (azs = $60^0$)

Como era de esperar, en ambos casos a menor valor de $E_s$, mayor valor de tasa de fotosíntesis (ver ecuación 1.7). También vemos que las condiciones para la máxima tasa de fotosíntesis se dan a profundidades de alrededor de 75 m para ángulo solar zenital de 0 grados; y a 50-60 m para 60 grados, pues en este último caso la menor presión ultravioleta permite a las células acercarse más a la superficie oceánica para captar más radiación fotosintéticamente activa.

Sin embargo, enfatizamos que las figuras anteriores pueden falsear nuestras conclusiones si ignoramos el papel de la circulación de Langmuir en el océano. Las figuras muestran las tasas de fotosíntesis como si los organismos unicelulares planctónicos estuvieran a profundidad fija todo el tiempo, pero en realidad esas células viviendo en la capa mezclada pueden ser fácilmente capturadas por las corrientes circulares de Langmuir, experimentando hundimiento y emergencia cíclicas.

A modo de ejemplo, consideramos esa circulación con una profundidad máxima de 40 metros, y calculamos las irradiancias promedio recibidas por un organismo atrapado en ésta usando las ecuaciones (1.8) y (1.9). Entonces sustituimos en la ecuación (1.4) para obtener la tasa fotosintética promedio en un ciclo completo. Los resultados de dicho procedimiento se muestran en la Tabla 1.2.

| $E_s$ (W.m$^{-2}$) | 15 | 20 | 25 |
|---|---|---|---|
| P/P$_s$ ( %) (azs = 0 deg) | 7,12 | 7,10 | 7,07 |
| P/P$_s$ ( %) (asz = 60 deg) | 15,8 | 15,2 | 14,5 |

Tabla1.2 Valores promedio, en un ciclo completo, para la tasa de fotosíntesis de una célula atrapada por una corriente circular de Langmuir.

Obtenemos pequeñas tasas promedios de fotosíntesis, de alrededor del 7%, para los tres valores de $E_s$ usados en este trabajo cuando el Sol está en el zenit, y valores de 15% para el ángulo solar zenital de 60 grados. Como podemos ver al comparar con las figuras 1.1 y 1.2, ser capturado en la circulación de Langmuir reduce considerablemente las posibilidades fotosintéticas de los organismos que viven en la capa mezclada, especialmente cuando hay una muy intensa irradiancia. Esto se debe a la cruda exposición a la radiación ultravioleta mientras el organismo está circulando en la parte superior del océano. Por debajo de la capa mezclada, la fotosíntesis sería razonablemente buena hasta 150 m y se extendería incluso a profundidades mayores que 200 m.

1.3.2 **Emergencia de vida en nuestra vecindad cósmica**

Es probable que la primera sonda no tripulada en probar a existencia de vida fotosintética en nuestra vecindad cósmica sea enviada a nuestro sistema planetario más cercano: el que pertenece a la estrella α de la constelación Centauro (α Centauri). Este sistema se encuentra a 4,37 años luz de nuestro Sol. Con la tecnología actual llegar allí tomaría varios milenios, pero no es descartable que tecnologías en desarrollo tales como la vela espacial o la fusión nuclear por pulsos podrían reducir este tiempo de manera considerable (Hearnshaw, 2010).

Realmente, α del Centauro es un sistema binario compuesto por dos estrellas bastante similares a nuestro Sol. α del Centauro A es una estrella G2V, como nuestro Sol, mientras que α del Centauro B es una estrella K1V. La masa de la estrella A es un 10% mayor que la del Sol mientras la masa de la B es alrededor de 10% menor. Las observaciones han descartado la existencia de planetas tipo Júpiter (gigantes gaseosos) y enanas pardas en este sistema, haciendo más probable la existencia de planetas tipo terrestre (Guedes y colaboradores, 2008). Sin embargo, éstos son más difíciles de detectar. Las simulaciones computacionales para la formación de planetas alrededor de α del Centauro B predicen de uno a cuatro planetas tipo terrestre en órbitas estables, uno o dos de ellos dentro de la zona estelar habitable (es decir, donde puede existir agua líquida en la superficie planetaria).

Utilizando las ecuaciones del epígrafe anterior, se puede hacer un cálculo aproximado para la tasa de fotosíntesis de un hipotético planeta tipo terrestre en la zona habitable de α del Centauro B (situado a 0,7 unidades astronómicas de esta estrella, lo cual asegura temperaturas superficiales similares a las de la Tierra). Usamos los datos de irradiancias ultravioletas para una estrella K2V dados en (Segura y colaboradores, 2003), como un modelo aproximado para α del Centauro B, que es una estrella K1V. Los resultados de este procedimiento se muestran en la figura 1.3.

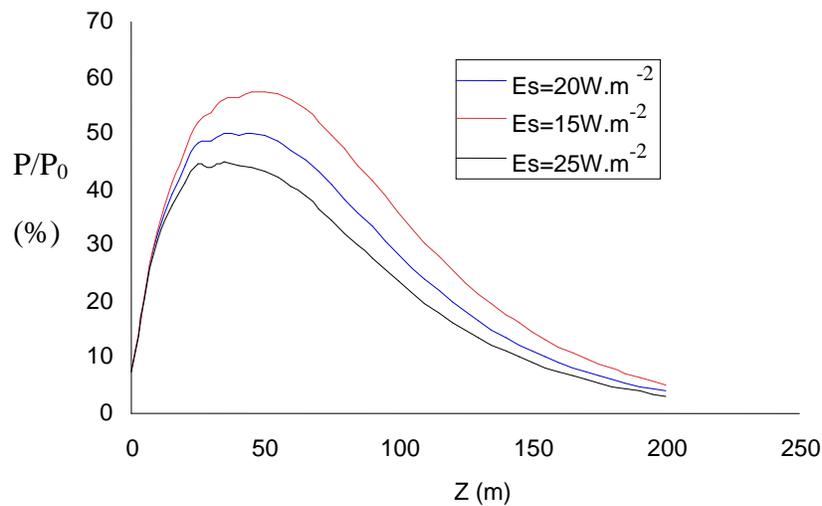

Figura 1.3 Tasa de fotosíntesis vs. Profundidad en el océano, para un ángulo solar zenital de 60 grados, en un planeta tipo terrestre a 0,7 UA de Alfa del Centauro B

Vemos entonces que las condiciones para la tasa fotosintética máxima están más cerca de la superficie si se compara con la Tierra Arcaica (compare con la Fig. 1.2). Esto se puede explicar porque Alfa del Centauro B es más fría que el Sol, emitiendo entonces una menor proporción de radiación ultravioleta. Al comparar las tasas máximas de fotosíntesis (Figs. 1.2 y 1.3), se infiere que hay condiciones similares o quizás, ligeramente más favorables para la emergencia de vida fotosintética en un planeta tipo terrestre localizado a 0,7 UA de Alfa del Centauro que en la Tierra Arcaica, si las demás condiciones ambientales son más o menos similares.

La situación hipotética descrita para (α Centauri) podría ser válida y extensible para otros sistemas más alejados dentro de nuestra galaxia. Si consideramos el número de sistemas estelares con características similares, incluso dentro de la llamada zona galáctica habitable, es posible estimar las posibilidades reales para el desarrollo de formas de vida fotosintéticas como un hecho nada despreciable.

### 1.4 La rotación planetaria y el régimen fotobiológico

Las condiciones para la vida en un planeta que orbita estrellas medianas (F, G, K) son radicalmente distintas a las de un planeta que orbita estrellas pequeñas (M). En este último caso el planeta deberá estar muy cerca de la estrella madre para que haya agua líquida en su superficie, pero entonces caerá en la zona de anclaje de marea y no podrá rotar sobre su propio eje, dando siempre la misma cara a la estrella. Esto implicará una ausencia de ciclicidad en el régimen fotobiológico; siempre habrá luz en el hemisferio que enfrenta la estrella y siempre oscuridad en el que queda detrás.

Por otro lado, para las estrellas medianas la zona habitable queda fuera de la zona de anclaje de marea, por lo que los planetas orbitantes podrán rotar sobre su propio eje y entonces habrá alternancia luz oscuridad. Es de esperar que esta ciclicidad del régimen fotobiológico impondrá ritmos circadianos (de cada día) en algunos de los procesos biológicos. En el caso de la Tierra, estos ritmos se observan en casi todos los organismos incluso en los seres humanos, siendo un ejemplo típico de este comportamiento la alternancia sueño vigilia. Investigaciones recientes (Uchida y colaboradores, 2010) muestran que estos ritmos alcanzan el nivel celular pudiendo estar representados tanto en la funcionabilidad de algunas estructuras foto-receptoras, como en el marco del propio código genético. Hoy se conoce que bajo determinadas condiciones los ritmos circadianos (originados por el régimen fotobiológico) modulan otros ritmos tales como el ciclo celular e incluso el ciclo de la glucólisis. La importancia biológica de los ritmos y particularmente del proceso de glucólisis se examinará en detalle en el capítulo III en un acercamiento a aquellas teorías que intentan explicar el origen de la vida.

1.5 **Conclusiones**

A lo largo de este capítulo hemos discutido varios aspectos relacionados directa o indirectamente con la emergencia de la vida y sus condicionamientos. En nuestro análisis hemos considerado elementos a escalas considerablemente diferentes que involucran aspectos importantes de la dinámica galáctica y cosmológica hasta particularizar en aquellos aspectos que rigen la vida fotosintética a escala planetaria en el marco de un sistema estelar determinado. Pese a las complejidades inherentes a un enfoque tan general, un análisis de esta naturaleza permite acercarnos desde una perspectiva polifacética, y por ende, más adecuada a la hora de abordar un fenómeno tan complejo y tan poco entendido como es la vida. Las conclusiones básicas derivadas de este capítulo podrían resumirse de la siguiente forma:

1. La velocidad de expansión del Universo ha permitido condiciones para la emergencia de vida, porque ha sido suficientemente moderada como para que se formen las estructuras necesarias (galaxias, estrellas, planetas).

2. La interacción entre los componentes mayoritarios energía y materia oscuras juegan un rol fundamental en el proceso de formación de estructuras a gran escala y por ende sobre el fenómeno de la Vida.

3. Existen mejores condiciones para la emergencia de vida en general, y fotosintética en particular, en los planetas que orbitan estrellas medianas (F, G y K), aunque no es de descartar la probable emergencia de vida en los sistemas de estrellas pequeñas (M).

4. Incluso para los escenarios más favorecidos dentro de la zona estelar habitable los niveles de radiación ultravioleta RUV conjuntamente con los efectos de la circulación vertical oceánica pueden limitar en buena medida las posibilidades reales de los organismos fotosintéticos dentro de la llamada zona fótica.

5. La temperatura de la estrella, la presencia o no de bloqueadores activos a la RUV tanto en la atmósfera como en el océano y las capacidades de reparación de la biota existente, emergen como parámetros claves a considerar en este tipo de estudios.

# LAS PERTURBACIONES DEL RÉGIMEN FOTOBIOLÓGICO EN LA VÍA LÁCTEA

## 2. LAS PERTURBACIONES DEL RÉGIMEN FOTOBIOLÓGICO EN LA VÍA LÁCTEA

El objetivo de este capítulo es examinar algunas perturbaciones en el régimen fotobiológico a que puede estar sometida la vida fotosintética en nuestra galaxia particularizando en el caso terrestre.

### 2.1 Explosiones estelares y la habitabilidad en la Vía Láctea

En todas las estrellas existe la posibilidad de emisión de grandes dosis de radiación ionizante al espacio. Los mecanismos son variados y con frecuencia implican reajustes en los campos magnéticos estelares, colapso gravitacional y/o establecimiento de reacciones nucleares muy energéticas.

Las explosiones estelares más energéticas que se conocen son los brotes de rayos gamma (BRG), en los que se emiten en escasos segundos o minutos energías del orden de $10^{44}$ J. Otro tipo de explosiones muy energéticas tienen lugar en las supernovas de diversos tipos. De hecho, existe una conexión directa entre algunos tipos de BRG y supernovas (Zhang y Meszaros, 2004). Estos conforman los llamados BRG de tipo II asociados al colapso de estrellas masivas (tipos O y B) como en el nacimiento de una supernova tipo 1b/c y se caracterizan por tener menor intensidad, mayor duración y una colimación más acentuada del haz. A diferencia de estos, los llamados BRG del tipo I se asocian a la fusión de sistemas binarios compactos como una estrella de neutrones y un agujero negro, o en los que una enana blanca incorpora materia de una gigante roja hasta alcanzar la masa crítica para que ocurra la explosión. Los brotes del tipo I se caracterizan por ser eventos muy energéticos, de menor duración así como un menor grado de colimación en el haz emergente si se comparan con sus homólogos del tipo II (Gao y Dai, 2009).

En el caso de las estrellas medianas y pequeñas, todavía en secuencia principal (o sea, que aún tienen hidrógeno como combustible nuclear), también hay tormentas de consideración que implican eyecciones de grandes cantidades de radiación al espacio. Por su relativa cercanía y elevada frecuencia, el efecto de estas tormentas puede llegar a ser importante para la vida en aquellos planetas tipo terrestres dentro la llamada zona estelar habitable.

Otros objetos compactos tales como las magnetoestrellas (un tipo de estrellas de neutrones, que por tanto no está en secuencia principal), también pueden liberar importantes dosis de radiación (Woods, 2004). Una de esas explosiones, provenientes de una distancia de 50 000 años luz, perturbó la ionosfera terrestre en el 2004 de manera medible. Alrededor de unas 12 magnetoestrellas han sido observadas en nuestra galaxia, pero podría haber más que aún no hemos detectado.

Se piensa que las explosiones estelares imponen restricciones a la habitabilidad de la Vía Láctea (González y colaboradores, 2001). Por ejemplo, no se espera vida en las cercanías del centro de la galaxia, pues la gran concentración de estrellas hace que la frecuencia de supernovas sea muy alta. Por otro lado, diversos autores han lanzado hipótesis sobre el probable rol de explosiones estelares en algunos descensos de biodiversidad en la Tierra (Smith y colaboradores, 2004b). Ejemplos son las hipótesis de que la extinción masiva del Ordovícico (Melott y colaboradores, 2004; Melott y Thomas, 2009), ocurrida hace unos 450 millones de años, fue causada por un BRG y que la extinción menor de moluscos bivalvos tropicales en la transición Pleistoceno-Plioceno fue causada por una supernova de la asociación Escorpión-Centauro (Benítez, Maíz-Apellániz y Canelles, 2002). En esta última asociación de estrellas masivas ocurrieron varias supernovas en los últimos 10 millones de años, y es intrigante la coincidencia de que las anomalías de $Fe^{60}$ en el lecho oceánico fijan como mejor fecha para una supernova cercana en unos 2,8 millones de años atrás, mientras que la extinción transcurre en varias fases, pero con un pico hace 2,6 millones de años. No obstante, se requiere de más estudio para demostrar fehacientemente a una supernova como la causa única, o más bien contribuyente junto a otras, de esta extinción.

El pasaje de un planeta por los brazos espirales de la galaxia también ha sido argumentado como relativamente peligroso para la vida tal y como la conocemos, debido en parte a la mayor cantidad de estrellas y por tanto de explosiones estelares. De hecho, algunos autores muestran que el Sistema Solar ha estado dentro de alguno de los brazos espirales durante casi todas las extinciones masivas del Fanerozoico (Leitch y Shavith, 1998).

Los principales efectos ambientales que una explosión estelar puede ocasionar sobre nuestro planeta (Scalo y Wheeler, 2002; Smith, Scalo & Wheeler, 2004a; Thomas y colaboradores, 2005) son espectros tipo aurora que depositan a nivel del mar un flash ultravioleta extremadamente intenso pero de corta duración (flash UV), formación de óxidos de nitrógeno que reducen la capa de ozono (con el consiguiente incremento del ultravioleta solar) y que bloquean parcialmente la luz visible con probable enfriamiento global. Otros de los efectos estimados son las lluvias ácidas, resultado de la combinación de los óxidos de nitrógeno con el agua atmosférica. Con la excepción del flash UV, los demás efectos pueden persistir por alrededor de una década, con importantes consecuencias para la biosfera, por supuesto en dependencia de la distancia a la que explota la estrella y a la energía liberada. Precisamente al cálculo de las distancias críticas para algunos de estos efectos dedicamos el próximo epígrafe.

2.2 **Principales efectos a corto plazo**

En este acápite estudiaremos los aspectos fundamentales asociados al flash UV y sus implicaciones potenciales sobre la biosfera.

2.2.1 **Interacción de los fotones gamma con las atmósferas planetarias**

Hemos seleccionado dentro de las explosiones estelares a aquellas que son las más energéticas, y por ende con mayor potencial para alterar considerablemente el régimen fotobiológico en un planeta tipo terrestre. Nos referimos a las explosiones o brotes de rayos gamma (BRG), en los que se emiten en escasos segundos o minutos energías del orden de $10^{44}$ J. Como ya hemos mencionado con anterioridad, la radiación de los BRG's sale colimada de la estrella progenitora bajo un ángulo relativamente variable en correspondencia con el tipo de evento. En este trabajo lo consideraremos aproximadamente de 0.01 srad en correspondencia con los trabajos de (Frail y colaboradores, 2001).

El proceso de interacción primario de los fotones gamma con la atmósfera desencadena toda una serie de fenómenos transitorios (ionización, excitaciones electrónicas, vibracionales, etc.) capaces de inducir incluso ionosferas secundarias y que culminan con la deposición progresiva de la energía original del haz en las diferentes capas de la atmósfera. Básicamente, el proceso de interacción para energías menores de 3 MeV se puede describir como la combinación de un proceso dispersivo de Compton y otro de foto-absorción, si tenemos en cuenta que otros procesos posibles, como la formación de pares, se encuentran considerablemente limitados a estas energías.

Por otro lado, si tenemos en cuenta que la energía de los fotones gamma supera considerablemente la energía de ionización típica de las moléculas, y que a su vez, estas últimas son aproximadamente similares para cualquier candidato atmosférico razonable (alrededor de unos 30-35 eV), es posible inferir que el proceso de interacción primaria con la atmósfera prácticamente no dependa de su composición: ellos ven un mar de electrones enlazados dentro de las moléculas y simplemente desprenden a muchos de ellos al interactuar con estas. Estos foto-electrones a su vez interactúan de diversas maneras excitando otras moléculas, las cuales al regresar a estados de menor energía emiten un espectro tipo aurora, en el que una fracción de la energía incidente llega al nivel del mar en forma de radiación ultravioleta: este es el vigoroso flash UV de unos 10 segundos de duración, primer efecto ambiental de un BRG. (Consultar Smith y colaboradores, 2004a para una descripción detallada de este proceso).

Para calcular la dependencia del efecto con la distancia a la estrella, primeramente asumimos un espectro del BRG (Smith y colaboradores, 2004a) la siguiente parametrización:

$$\frac{dN}{dE} = k\left(\frac{E}{100 keV}\right)^{\alpha} \exp\left(-E/E_0\right) \quad \text{para} \quad E \leq (\alpha - \beta)E_0 \qquad (2.1)$$

$$\frac{dN}{dE} = k\left[\frac{(\alpha - \beta)E_0}{100 keV}\right]^{\alpha - \beta} e^{\beta - \alpha}\left(\frac{E}{100 keV}\right)^{\beta} \quad \text{para} \quad E \geq (\alpha - \beta)E_0, \quad (2.2)$$

donde $E_0 = 250 keV$, $\alpha = -0.9$ y $\beta = -2.3$.

El rango de energías para los fotones gamma de 50 keV < E < 3 MeV. Este espectro original fue dividido en 100 intervalos "monocromáticos" igualmente espaciados logarítmicamente, con un valor correspondiente de energía media y flujo asociado a cada intervalo.

Para todos los modelos atmosféricos se consideró una atmósfera exponencial y con constituyentes bien mezclados (excepto el ozono), con una escala de altura Ho = 8 km y una densidad $\rho = 1,3 \times 10^{-3}$ g/cm$^3$.

Para la deposición de energía gamma (liberación de electrones y posterior excitación de las moléculas), se asumió la composición actual de la atmósfera, pues como dijimos anteriormente estos procesos poco dependen de la composición. Entonces se empleó el procedimiento general descrito en (Gehrels y colaboradores, 2003) . La atmósfera se dividió en 90 capas desde la altura 0 (nivel del mar) hasta 180 km, y el rango de longitudes de ondas se dividió en 100 haces "monocromáticos". Cada haz normalmente incidente fue atenuado por una ley exponencial de Beer de la forma $N_{i,j} = N_i^0 e^{-\mu_i x_j}$, de acuerdo a su propio flujo y energía media, donde $N_{i,j}$ es el flujo de fotones remanente del haz monoenergético incidente $N_i^0$; $x_j$ es la densidad de columna (en g/cm$^2$) de la capa y $\mu_I$ es el coeficiente de atenuación másico correspondiente, tomado de (Berger y colaboradores, 2005). La energía depositada en cada capa fue determinada usando la diferencia de los flujos totales de capas adyacentes:

$$Fdep, i = \sum_{j=1}^{100} \left( N_{i,j} - N_{i,j-1} \right) \quad (2.3)$$

El proceso de posterior emisión del espectro tipo aurora (que incluye el flash UV) sí depende de la composición atmosférica, por lo que se tomaron cinco modelos de paleo-atmósferas; con $10^{-5}$, $10^{-4}$, $10^{-3}$, $10^{-2}$ y $10^{-1}$ naa (nivel atmosférico actual) de O$_2$, respectivamente. También se tomó la atmósfera moderna con $10^0$ naa. Podemos considerar que a grandes rasgos nuestros modelos de atmósferas coinciden con razonable aproximación a aquellas existentes en los tiempos geológicos que se muestran en la tabla 2.1.

| Contenido de oxígeno en nivel atmosférico actual ($O_2$ naa) | Tiempo geológico |
| --- | --- |
| 1 | Proterozoico Tardío y eón Fanerozoico (0.8 Ga atrás hasta el presente) |
| $10^{-1}$ or $10^{-2}$ | Proterozoico Medio (2.3 Ga – 0.8 Ga antes del presente) |
| $10^{-3}$, $10^{-4}$ or $10^{-5}$ | Eón Arcaico y Proterozoico Temprano (3.8 Ga – 2.5 Ga) |

Tabla 2.1 Atmósferas modelo estudiadas en el tiempo geológico

En los demás aspectos seguimos a (Segura y colaboradores, 2003): adoptamos sus datos para la columna de ozono, fijamos la presión atmosférica a 1 atm y en las paleo-atmósferas reemplazamos el $O_2$ faltante por $CO_2$, al ser este un gas que probablemente abundó en los paleo-ambientes de nuestro planeta.

Para implementar la columna de ozono en cada una de las atmósferas se siguieron dos procedimientos. El primero fue un simple re-escalado de la atmósfera estándar USA 76, y el segundo implicó una bajada en altura de todo el perfil del ozono y un posterior re-escalamiento al caso apropiado. Se conoce que la depresión del oxígeno no sólo hace la capa de ozono más delgada, sino más cercana a la troposfera (Segura y colaboradores, 2003; Kasting y Catling, 2003). El segundo procedimiento intenta mimetizar a grandes rasgos el último efecto.

Considerando lo anterior, la reemisión de la parte ultravioleta del espectro tipo aurora se asume que sigue la ley empírica (Smith y col 2004):

$$\frac{dF_{UV,i}}{d\lambda} = \frac{Fdep,i}{\lambda \ln\left(\lambda_{max}/\lambda_{min}\right)} \qquad (2.4)$$

donde $Fdep,i$ es la energía total depositada en la capa i-ésima. Los valores $\lambda_{max}$ y $\lambda_{min}$ se asocian a los límites típicos de las bandas del nitrógeno y se fijaron a 130 y 600 nm respectivamente. Si tenemos en cuenta que el nitrógeno es actualmente un componente mayoritario en la atmósfera y que al parecer, su columna no ha variado notablemente durante la evolución geológica de la Tierra, tal elección parece justificada. De cualquier manera, probablemente la expresión 2.4 sea ajustable para describir, al menos groseramente, el comportamiento espectral de otras especies moleculares importantes. Finalmente, para obtener los flujos de radiación UV en la superficie de la Tierra, se tuvieron en cuenta los efectos de la absorción y de la dispersión de Rayleigh para cada uno de los modelos de paleo-atmósferas propuestos. El efecto neto en la superficie se calcula como la suma de los efectos de emisión de cada capa.

La tabla 2.2 muestra la energía UV que alcanza el suelo, expresada como fracción de la energía gamma originalmente incidente en el tope de la atmósfera para algunas bandas representativas, mientras en la Fig. 2.1 se muestra los espectros UV en función de la longitud de onda λ que alcanzan la superficie terrestre para los distintos modelos convenientemente normalizados respecto al flujo total que deposita el BRG en el tope de la atmósfera. De análisis de ambos se puede apreciar que incluso llegan fotones de la banda UV-C al suelo, o sea, con longitudes de onda por debajo de 280 nm, lo cual no sucede típicamente con la irradiancia solar por la atenuación atmosférica. La banda UV-C es particularmente dañina desde el punto de vista biológico entre otros factores por contener el pico de máxima absorción del ADN y por la marcada formación de radicales libres.

| $O_2$ naa | UV-A<br>315-400 nm | UV-B<br>280-315 nm | UV-C<br>130-280 nm | TOTAL<br>130-600 nm |
|---|---|---|---|---|
| $10^0$    | $2.330 \times 10^{-02}$ | $6.311 \times 10^{-04}$ | $3.043 \times 10^{-12}$ | $2.022 \times 10^{-01}$ |
| $10^{-1}$ | $5.867 \times 10^{-02}$ | $2.789 \times 10^{-03}$ | $2.418 \times 10^{-09}$ | $2.480 \times 10^{-01}$ |
| $10^{-2}$ | $6.484 \times 10^{-02}$ | $5.161 \times 10^{-03}$ | $4.817 \times 10^{-07}$ | $2.583 \times 10^{-01}$ |
| $10^{-3}$ | $6.598 \times 10^{-02}$ | $8.943 \times 10^{-03}$ | $5.802 \times 10^{-05}$ | $2.643 \times 10^{-01}$ |
| $10^{-4}$ | $6.616 \times 10^{-02}$ | $1.187 \times 10^{-02}$ | $2.338 \times 10^{-03}$ | $2.699 \times 10^{-01}$ |
| $10^{-5}$ | $6.617 \times 10^{-02}$ | $1.210 \times 10^{-02}$ | $2.933 \times 10^{-03}$ | $2.707 \times 10^{-01}$ |

Tabla 2.2 Energía UV que alcanza el suelo expresada como fracción de la energía gamma originalmente incidente en el tope de la atmósfera.

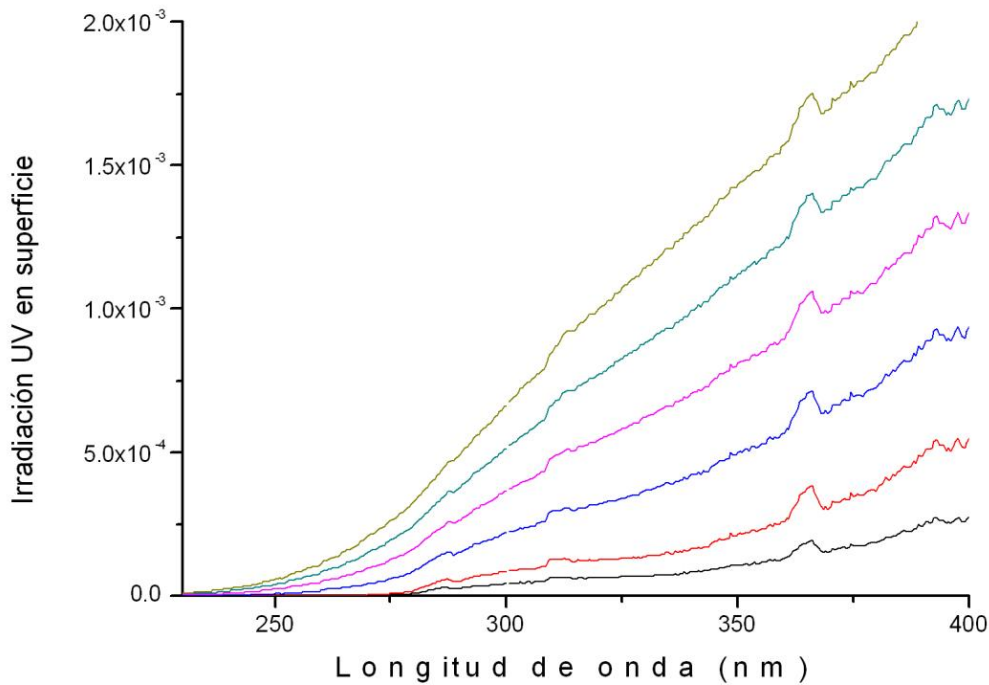

Fig. 2.1 Irradiancias UV que alcanzan el suelo para distintas atmósferas y longitudes de onda, normalizadas respecto al flujo completo del BRG en el tope de la atmósfera.

2.2.2 **Efectos biológicos y distancias críticas**

A partir de los resultados derivados en la sección anterior es posible estimar a que distancias el progenitor de un BRG, con las características antes mencionadas, pueda considerarse potencialmente peligroso para la vida en nuestro planeta. Para calcular estas distancias críticas recurriremos a algunos criterios e índices reportados en la literatura sobre el tema así como a estudios que evidencian el impacto negativo que tienen incluso los niveles moderados de RUV solar sobre una buena parte de los productores primarios.

Con ese fin y similar al tratamiento realizado en el primer capítulo para el eón arcaico, calculemos las irradiancias biológicas efectivas E* asociadas al BRG para cada modelo de paleo-atmósfera considerado. Para ello convolucionaremos el espectro de acción biológica e(λ) para daño al ADN con la irradiancia espectral UV incidente E(λ) según la expresión (Cockell 2000; Cockell y Raven 2007).

$$E^* = \sum e(\lambda)E(\lambda)\Delta\lambda \qquad (2.5)$$

Definamos entonces las fluencias biológicas efectivas F* como:

$$F^* = E^* \Delta t, \qquad (2.6)$$

donde Δt es el tiempo de exposición a la radiación UV. Esta magnitud estima el efecto de la radiación integrado en el tiempo y suele ser conveniente en aquellos casos donde la contribución de los mecanismos intrínsecos de reparación de daños (por ejemplo, los mecanismos reparadores del ADN) sea poco significativa. De esto no ser cierto, el daño biológico estimado directamente a partir de la magnitud $F^*$ podría estar seriamente sobredimensionado. Conviene señalar en este punto que tanto para el caso particular de $F^*$ como de cualquier otra expresión matemática que se introduzca en lo que sigue, el subíndice BRG denotará siempre el UV producto de la explosión estelar, mientras que el subíndice Sol denotará el UV proveniente del Sol.

En lo que respecta a la resistencia a las radiaciones, la enorme variabilidad encontrada entre las especies hace que definir un criterio de daño biológico significativo se considere un problema borroso. Con vistas a solventar en alguna medida estas dificultades consideraremos, al menos por el momento, valorar la acción en los productores primarios más abundantes, que son los organismos fotosintéticos unicelulares: el fitoplancton. Cualquier afectación importante sobre el fitoplancton repercutirá sobre gran parte de la biosfera (Behrenfeld y colaboradores, 2005) por su transmisión a través del ensamblaje trófico (alimentario) y mediante otros mecanismos indirectos. En nuestro caso modificamos el criterio en (Thomas y colaboradores, 2005), pues en vez irradiancias o flujos biológicos efectivos, introducimos el concepto de fluencias biológicas efectivas para describir el potencial daño del BRG. La razón para esta elección está en que durante el corto e intenso flash UV de 10 s de duración no habrá tiempo suficiente para que los mecanismos de reparación del ADN funcionen, y por ende los daños serán acumulativos, propiamente descritos por una fluencia (J/m$^2$) y no por un criterio de flujo o irradiancia (W/m$^2$), más adecuado para escenarios en los que la reparación del ADN funciona. Basados en esto, asumimos la siguiente condición para tener una significativa mortalidad de fitoplancton superficial bajo la acción del UV flash:

$$F^*_{BRG} = nF^*_{Sol} \qquad (2.7)$$

Esto significa que durante el tiempo de exposición de 10 s al flash UV, las células de fitoplancton en la superficie acuática estarían expuestas a una fluencia biológica efectiva n veces mayor que la que están acostumbradas a recibir desde el Sol en un día entero. Parece razonable esperar mortalidad significativa incluso para n=1, especialmente considerando que durante los 10 s del UV flash no habrá tiempo para los mecanismos de reparación del ADN. Por supuesto, la mortalidad no solo ocurrirá durante el breve tiempo de exposición, sino también con posterioridad a medida que las secuelas irreversibles inducidas sobre los organismos se vayan evidenciando. Otro elemento que refuerza esta elección son los estudios recientes que muestran que una buena parte de los productores primarios, especialmente el llamado pico-fitoplancton, recibe dosis de UV solar significativas durante el día, responsables de la elevada tasa de mortalidad reportada para estos organismos. O sea, incluso bajo las condiciones relativamente suaves de irradiación de la era moderna, al parecer un número considerable de especies que habita nuestro planeta se encuentran notablemente estresadas por efecto del UV solar.

Otra cuestión a considerar es la duración del día en los diferentes eones. Aun cuando hay incertidumbres, una duración de 15 horas en el Arcaico temprano parece bien fundamentada, así como una de 20 horas en el Proterozoico, la que ha ido aumentando hasta las 24 horas actuales. Tomamos un valor promedio de 20 horas y suponemos un fotoperiodo promedio de 10 horas para la exposición diaria real al Sol. Las diferencias en cuanto a eón y latitud geográfica no cambian apreciablemente nuestros resultados, pues se trata de modelaciones a escala planetaria y no regional.

Definamos ahora la irradiancia biológica efectiva en una forma adimensional según la expresión:

$$E_{ad}^{*} = \frac{E_{BRG}^{*}}{I_{BRG}^{*}}, \qquad (2.8)$$

donde $I_{BRG}^{top}$ es el flujo total del BRG en el tope de la atmósfera. Si graficamos entonces esta magnitud para cada modelo de atmósfera considerado obtenemos la figura 2.2

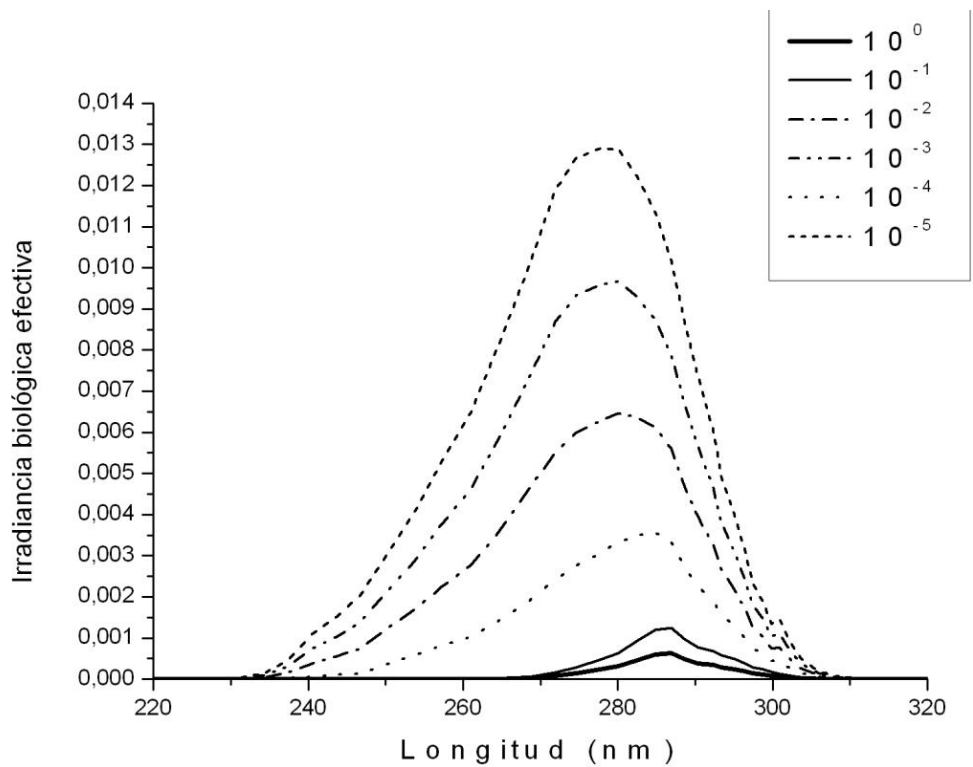

Figura 2.2 Irradiancias biológicas efectivas normalizadas con respecto al flujo total del BRG en el tope de cada atmósfera estudiada.

Como era de esperar, cuanto menor sea el contenido de oxígeno, mayor será el daño en el ADN causado por el BRG, ya que la atmósfera tendría menos ozono para proteger los organismos del UV retransmitido. Los resultados graficados en la figura 2.2 constituyen un primer paso para explorar la importancia relativa del flash. Tengamos en cuenta que por lo general, los autores que trabajaron en la modelización de los efectos de la GRB en la Tierra tienden a ignorar el flash UV, en gran parte basado en el argumento de que los efectos a largo plazo son más importantes. Además, en ambientes que poseen un escudo protector de oxígeno-ozono, la banda UV-C es absorbida casi en su totalidad, y su efecto biológico eficaz es relativamente pequeño, como puede verse en las dos curvas inferiores de la Fig. 2.2. Sin embargo, con la disminución progresiva del contenido de oxígeno, aparece una fracción notable en la región del UV-C que alcanza la superficie del planeta y que contribuye de manera muy negativa, no solo sobre el ADN, sino sobre la mayoría de los procesos biológicos. Por lo tanto, para las atmósferas que tienen de $10^{-3}$ a $10^{-5}$ naa (probablemente durante el Arcaico y Proterozoico temprano) el pico de la irradiancia biológicamente efectiva cae en la región UV-C, como puede verse en las tres curvas de la parte superior de la Fig. 2.2. Esto indica, en nuestra opinión, que valdría la pena investigar los posibles efectos biológicos del flash UV, especialmente para ambientes con bajos valores de la columna de ozono.

Ahora, si combinamos las ecuaciones (2.6-2.8) obtenemos:

$$E_{ad}^{*} I_{BRG}^{top} \Delta t_{BRG} = n E_{Sol}^{*} \Delta t_{Sol} \qquad (2.9)$$

Si tenemos en cuenta que la fluencia del BRG en el tope de la atmosfera se puede escribir como:

$$I_{BRG}^{top} \Delta t_{BRG} = \frac{E_{\gamma}}{D^2 \Delta \Omega} , \qquad (2.10)$$

donde $E_{\gamma}$=5x10$^{43}$ J es la energía gamma, $\Delta\gamma$ es el ángulo sólido, fijado a 0,01 srad de acuerdo a (Frail y colaboradores, 2001) y D es la distancia a la estrella. La absorción en el medio interestelar no se incluyó en nuestro análisis debido a ser este prácticamente transparente a los fotones de alta energía (Galante y Horvath, 2007).

Sustituyendo la ecuación (2.10) en la (2.9) y resolviendo para D obtenemos la expresión para calcular las distancias críticas a partir de la cual una estrella emisora de un BRG causaría mortalidad superficial significativa al cumplimentarse la condición (2.7):

$$D = \sqrt{\frac{E_\gamma \times E_{ad}^*}{\Delta\Omega \times \Delta t_{Sol} \times n \times E_{Sol}^*}} \qquad (2.11)$$

Para estimar $E_{Sol}^*$ para nuestras atmósferas proponemos el siguiente procedimiento aproximado: consideramos que la atmósfera Arcaica estudiada en (Cockell, 2000) se corresponde a nuestro modelo de atmósfera con $10^{-5}$ naa $O_2$. De la Fig. 5 de dicho artículo se puede inferir valores aproximados de 80 W/m$^2$ y 20 W/m$^2$ para ángulos zenitales solares de 0 y 60 grados, respectivamente. Como un promedio aproximado usamos $E_{Sol}^* = 50 W/m^2$. Seguidamente utilizamos la Tabla 3 de (Segura y colaboradores, 2003), donde los $E_{Sol}^*$ para las mismas paleo-atmosferas aparecen normalizados al valor actual, lo cual nos permite estimar $E_{Sol}^*$ para todas nuestras atmósferas. En el mencionado artículo los autores usan el rango 295-315 nm como un estimador para todos los efectos del UV. Esto implicara alguna subestimación en $E_{Sol}^*$ debido a la absorción de la banda UV-C (200-280nm) por el ozono. Por tanto, a menor contenido de oxígeno, dicha subestimación será mayor. Sin embargo, consideramos que los errores introducidos serán relativamente pequeños, especialmente si se tiene en cuenta que se está trabajando con valores relativos (normalizados) para inferir los $E_{Sol}^*$ correspondientes a cada atmósfera.

En la Tabla 2.3 a continuación se resumen los principales resultados derivados del esquema anterior y al empleo de la expresión (2.11).

La magnitud de las distancias reportadas en la tabla anterior posibilitan hacer algunas conclusiones generales sobre la posibilidad y magnitud que un evento de este tipo pudo haber tenid en algún momento de la evolución biológica de nuestro planeta y por ende aplicables a cualquier otro planeta similar al nuestro en la galaxia. Destacamos las siguientes:

| $O_2$ naa | $E_{ad}^{*}$ | $E_{Sol}^{*}(W/m)$ | $D_2(kpc)$ | $D_4(kpc)$ | $D_8(kpc)$ | $D_{16}(kpc)$ |
|---|---|---|---|---|---|---|
| $10^{0}$ | 0.005 | 0.009 | 4.45 | 3.14 | 2.22 | 1.57 |
| $10^{-1}$ | 0.010 | 0.0141 | 5.07 | 3.58 | 2.53 | 1.79 |
| $10^{-2}$ | 0.043 | 0.095 | 4.00 | 2.83 | 2.00 | 1.42 |
| $10^{-3}$ | 0.085 | 1.961 | 1.25 | 0.88 | 0.63 | 0.44 |
| $10^{-4}$ | 0.128 | 33.454 | 0.37 | 0.26 | 0.18 | 0.13 |
| $10^{-5}$ | 0.172 | 50 | 0.35 | 0.25 | 0.17 | 0.12 |

Tabla 2.3 Distancias críticas para que el flash UV que llega a nivel del mar supere un número determinado de veces la fluencia biológica efectiva del UV solar.

Como primer elemento tenemos que, las biosferas del Fanerozoico, del meso-Proterozoico y Proterozoico tardío serían, a corto plazo, las más estresadas por un flash de UV de un GRB procedente de varios *kpc*. Para esto tengamos en cuenta que incluso, para causar un daño 16 veces mayor que el asociado al UV solar (última columna), las distancias críticas para esas épocas son similares a las estimadas para el "último estallido cercano típico" a la Tierra, o sea de 1 a 2 kpc en los últimos 1000 millones de años. Por lo tanto, podría ser interesante la inclusión del flash UV dentro del modelado detallado de los efectos a largo plazo de un BRG durante el eón Fanerozoico, tal como se presenta en Thomas y colaboradores del 2005.

Por otro lado, si nos enmarcamos en una distancia dada al progenitor del BRG, los ecosistemas que viven en atmósferas del orden de unos $10^{-1}$ naa serían entonces los más estresados. Este hecho corresponde aproximadamente a una atmósfera anterior a la mitad del período Proterozoico.

Se desprende además un hecho que a primera vista puede parecer sorprendente. Los ecosistemas menos protegidos, en correspondencia con el análisis las tres últimas filas de la Tabla 2.3, no son necesariamente los más afectados. Ciertamente, recibirían una mayor fluencia biológicamente efectiva de un BRG, pero sus efectos serían menos significativos al recibir mayores fluencias biológicamente efectivas durante el día. Por lo tanto, es atractivo pensar que durante el Proterozoico y el Arcaico temprano la biota más resistente a las radiaciones residía en tierra firme. Por lo tanto, un BRG la hubiera podido afectar menos que a la biota posterior que solía vivir en entornos fotobiológicamente más amenos, tales como los de la mitad del Proterozoico hasta el presente.

## 2.3 Principales efectos a largo plazo

### 2.3.1 Incrementos persistentes de los niveles del UV solar asociados a un BRG.

Se conoce de las secciones anteriores que un BRG puede afectar los niveles típicos de radiación UV que alcanzan el suelo en al menos dos formas diferentes, el ya estudiado flash UV y por el aumento del UV solar asociados a la disminución de la capa de ozono (Galante y Horvath, 2007).

La importancia relativa de estos efectos parece ser una función fuerte del contenido de oxígeno libre en la atmósfera.

Para atmósferas contemporáneas, similares a la de la Tierra (rica en $O_2$), la principal influencia estimada del BRG es el de la disminución de la capa de ozono y el consecuente incremento del UV solar en superficie, siendo ambos efectos consecuencia directa de la formación de grandes cantidades de óxidos de nitrógeno. La recuperación total de la capa de ozono estará determinada principalmente por fenómenos de transporte con una escala de tiempo estimada de alrededor de una década (Thomas y colaboradores, 2005). Según este estudio, el llamado "destello típico más cercano" que hipotéticamente ocurrió en los últimos 1000 millones de años podría causar, a nivel mundial, una disminución en la capa de ozono de hasta un 38% y una fracción importante de esta (por lo menos 10%) se mantendría hasta por siete años implicando:

- Una mayor irradiación de la superficie del planeta con la radiación solar ultravioleta (RUV),

- Aumento de la opacidad de la atmósfera reduciendo la luz del sol visible en unos pocos %, debido a la formación de NO2, con potencial enfriamiento global (Melott y colaboradores, 2005)

- Depósito de nitratos por lluvias ácidas en cantidades algo mayores que las causadas por los rayos cósmicos en la actualidad.

Es evidente que los dos primeros efectos podrían afectar a muchas especies fotosintéticas: más UV solar puede dañar el ADN e inhibir la fotosíntesis hasta cierto punto, mientras que la reducción de luz del sol en el visible (es decir, la radiación activa fotosintéticamente activa, RFA) podría reducir la energía disponible para realizar la fotosíntesis y por lo tanto para la producción primaria. Sin embargo, el tercer efecto puede compensar, al menos parcialmente, la mencionada inhibición de la fotosíntesis, y podría incluso causar la eutrofización (enriquecimiento de nutrientes en exceso) en algunos ecosistemas de agua dulce y costeros. Es cierto que la lluvia ácida podría tensionar sectores de la biosfera, pero después de una titulación, el aumento del nitrato depositado podría ser útil para los organismos fotosintéticos, especialmente para las plantas terrestres.

### 2.3.2 Indicadores globales de daño biológico

Con vistas a estudiar los efectos del flash UV, en las secciones anteriores habíamos considerado varios criterios cuantitativos para evaluar los daños provocados por la radiación. Aunque estos criterios son igualmente aplicables al incremento persistente de la RUV solar, es conveniente en cualquier caso, estudiar la posibilidad de definir algunos criterios de daños que nos permitan tener una perspectiva global de esta influencia sobre la biosfera. Sobra reiterar las múltiples dificultades y factores que conspiran contra establecer un criterio de esta naturaleza. Sin embargo, una idea aproximada de estos efectos desde una perspectiva global es dada por el llamado factor de amplificación de la radiación (FAR), relacionando las irradiaciones biológicamente efectivas $E^*$ (solares) con las columnas de ozono $N$, antes y después del evento ionizante.

$$\frac{E^*_{después}}{E^*_{antes}} = \left(\frac{N_{antes}}{N_{después}}\right)^{FAR} \qquad (4)$$

Los valores de lãs *FAR* son dependientes tanto del grupo de especies como del proceso orgánico a considerar (representada por una función de ponderación biológica o *FPB*). Dichas funciones son típicamente medidas en condiciones controladas de laboratorio, de modo que tienen un valor limitado para estimar la respuesta real de los seres vivos a la RUV.

Por otro lado y aunque no han sido considerado explícitamente, bajo la acción de la RUV los organismos pueden revertir las reacciones fotoquímicas utilizando enzimas especializadas o volver a sintetizar las moléculas afectadas. Estos procesos, conocidos genéricamente como de reparación, no dependen sólo de la especie, sino también de las variables ambientales. Por ejemplo, consideremos la relación de reparación-temperatura conocida para varias especies de fitoplancton: a muy bajas temperaturas la reparación es muy lenta, mientras que a temperaturas más altas la reparación es eficiente. Otro factor climático a considerar es la propia irradiación luminosa, fundamentalmente en la llamada banda de foto-reactivación sobre los (350-450 nm) que incluye la luz azul y parte de la banda UV-A. La importancia de esta banda se debe a que la fotoliasa, enzima que garantiza uno de los mecanismos de reparación más eficientes y extendidos entre los organismos, precisa de este rango para activarse. En una reacción química ultra-rápida, dicha enzima es capaz de eliminar lesiones como los dímeros de timina, una de las más mutagénicas y tóxicas inducidas por la radiación ultravioleta al ADN (para más detalles consultar Sinha y Hader, 2002). Adicionalmente a la foto-reactivación que utiliza una única enzima, la célula dispone de varias estrategias y mecanismos de reparación en cadena que involucran a múltiples enzimas y que no precisan de la luz visible por lo que suelen denominarse genéricamente como reparación en oscuro.

Como en general, dichos mecanismos no se tiene en cuenta a la hora de medir las FPB, es el factor de amplificación biológico (FAB) la cantidad que nos da una información más precisa sobre los efectos biológicos de la RUV:

$$\frac{P_{después}}{P_{antes}} = FAB \times \frac{E^*_{después}}{E^*_{antes}} \qquad (2.5)$$

donde $P$ es la tasa de un proceso orgánico (por ejemplo, fotosíntesis). Aunque muy pocos *FAB* se han medido experimentalmente, conjuntamente con la FAR, estos podrían ser útiles para una primera estimación del daño global en la biosfera por un BRG, aunque una modelización más detallada implica que uno debe estudiar ecosistemas específicos, las unidades básicas para la construcción de la biosfera.

Los resultados de aplicar estos criterios al caso de un BRG típico descrito en Thomas y colaboradores, (2005) se recogen en la Tabla 2.4, mostrándose el aumento de la fracción efectiva de irradiación biológica para diferentes valores de la disminución del ozono y varias funciones de ponderación biológica.

| **Funciones de peso biológico** | **RAF** | $\dfrac{E^*_{después}}{E^*_{antes}}$ para disminuciones del ozono en  (%) | | | |
|---|---|---|---|---|---|
| | | **38** | **30** | **20** | **10** |
| **Foto-inhibición del fitoplancton marino** | **0.31** | 0.31 | 1.12 | 1.07 | 1.03 |
| **Foto-inhibición en plantas terrestres** | **0.51** | 0.51 | 1.20 | 1.12 | 1.05 |
| **Daño al ADN** | **1.67-2.2** | 2.22-2.85 | 1.82-2.20 | 1.45-1.63 | 1.19-1.26 |

Tabla 2.4 Factores de amplificación de la radiación y el incremento fraccional de las irradiancias biológicas efectivas para reducciones de la columna de ozono de 38, 30, 20, 10 porciento.

Dicha tabla sugiere que el daño al ADN es, en general, la principal influencia de un BRG sobre la biosfera y que las plantas terrestres pueden sufrir incluso más que el fitoplancton. Sin embargo, debe tenerse en cuenta que los FAR se miden en condiciones de control muy diferentes a las condiciones naturales en las cuales viven los organismos. Por lo tanto, el uso de factores de amplificación biológicos (FAB) o curvas de respuesta a la exposición (CRE) debe darnos una idea mucho mejor de la respuesta de la biosfera a las perturbaciones de la RUV. Desafortunadamente, muy pocos FAB o CRE se han medido para los productores primarios más comunes en la biosfera, tales como las principales especies de fitoplancton marino. Por lo tanto, nos faltan datos biológicos de campo para efectuar cálculos más precisos de los posibles efectos globales de un BRG en la biosfera. Por suerte varios estudios están en marcha que proporcionarán datos biológicos de utilidad, por lo que el futuro próximo parece bastante prometedor.

### 2.3.3 Reducción de los niveles de irradiación en superficie asociados al $NO_2$

Para tener en cuenta la reducción del espectro de la irradiación en la superficie del planeta debido a la formación de $NO_2$, se utilizó el espectro de energía solar $I_0(\lambda)$ en la superficie que figura en la norma ASTM G173-03e1 (Http://www.astm.org/Standards/G173.htm). Así, teniendo en cuenta que en (Thomas y colaboradores, 2005) se reporta una reducción de la irradiación total en el rango (0-10)% debido a la formación $NO_2$, calculamos el valor de las columnas de $NO_2$ que llevarían a una reducción de la irradiación total $I$ para varios valores de la fracción $f$ según:

$$\frac{I_{después}}{I_{antes}} = f \qquad (2.6)$$

donde a partir de ahora los subíndices *antes* y *después* denotan los valores posterior y anterior del impacto del BRG. Se utilizaron los valores de $f$ de 0.98, 0.96, 0.94 y 0.92, los cuales representan reducciones en la irradiación de 2, 4, 6 y 8% respectivamente. Los valores de la irradiación total antes y después del BGR están dados por

$$I_{antes} = \int_{280nm}^{700nm} I_0(\lambda)\, d\lambda \qquad (2.7)$$

$$I_{después} = \int_{280nm}^{700nm} I_0(\lambda)\, e^{-\tau} d\lambda \qquad (2.8)$$

donde τ es el camino óptico de los fotones en la columna de $NO_2$. Esta magnitud da la clave para estimar la cantidad de $NO_2$ necesaria para reducir la irradiación total en un factor *f* dado.

Los resultados de aplicar dicha metodología se muestran en la tabla 2.5 para algunas bandas de interés seleccionadas. En ella se muestra la presencia de una tenue banda de absorción en la banda visible (RFA), mientras que una absorción más pronunciada aparece en la región del UV-A y en la banda de foto-reparación biológica (350-450 nm). Los efectos sobre esta última puede ser importantes si tenemos en cuenta que la luz en este rango (azulada) como ya se discutió es necesaria para ejecutar la foto-reparación (Sinha y Hader, 2002).

| *f* | *f* UV-A | *f* PAR | *f* 350–450 nm |
|---|---|---|---|
| 0.98 | 0.92 | 0.98 | 0.92 |
| 0.96 | 0.85 | 0.95 | 0.84 |
| 0.94 | 0.78 | 0.93 | 0.77 |
| 0.92 | 0.71 | 0.90 | 0.70 |

Tabla2.5 Cocientes de irradiación antes y después del BRG para algunas bandas seleccionadas.

El procedimiento anterior no considera el aumento de la irradiación debido a la reducción del ozono, pero como el Sol tiene su máximo en la parte visible del espectro, la contribución a la irradiancia total es muy pequeña. Esta afirmación fue confirmada mediante el uso del código de transporte radiativo NCAR / ACD TUV: Tropospheric Ultraviolet & Visible Radiation Model (http://cprm.acd.ucar.edu/Models/TUV/). Valiéndonos del citado código pudimos comprobar que un 30% de disminución de la columna de ozono standard de 340 unidades Dobson implicó un aumento del 22% de la radiación UV-B, pero un incremento de sólo 0,37% en la región del UV-A. Este resultado confirma que la contribución al aumento de la radiación UV-A asociado al descenso del ozono resulta prácticamente despreciable frente al efecto de disminución asociado a la formación de $NO_2$, que en este caso está en torno del 10%.

Resumiendo los resultados anteriores, podemos apreciar que el efecto global neto de un BRG en atmósferas contemporáneas (ricas en $O_2$) sugiere una combinación de varios tipos de daños. Básicamente tendríamos un incremento de la R UV (disminución de la capa de ozono), y una eficiencia menor en los mecanismos de reparación del daño al ADN. Además, las reducciones en el visible implican que menos luz (RFA) estará disponible para la fotosíntesis. Por último, la reducción total de la luz solar en los porcentajes calculados en este trabajo podría indicar un enfriamiento global, hecho que de por sí merece consideración en futuras investigaciones.

2.4 **Conclusiones**

Durante el transcurso de este capítulo se han discutido varios aspectos sobre las posibles influencias que una explosión estelar tan intensa como un BRG puede tener sobre la vida en nuestro planeta y particularmente sobre los productores primarios más extendidos. Hemos empleado, discutido y definido criterios que nos permiten estimar la importancia relativa de algunos efectos asociados a este tipo de eventos como son el flash UV, los indicadores de daño global o el incremento de la opacidad asociada al $NO_2$. Parte de los resultados fundamentales, como los derivados en la sección 2.3 (efectos a largo plazo), son aplicables también en el contexto de otros eventos estelares como pueden ser las explosiones de supernovas. Básicamente las conclusiones fundamentales derivadas en el capitulo podrían resumirse así:

1. Perturbaciones intensas del régimen fotobiológico asociadas a eventos astrofísicos como un BRG pueden potencialmente afectar la biosfera en correspondencia con la distancia al progenitor y al nivel de oxígeno atmosférico libre.
2. Los ecosistemas que habitan en atmósferas con contenido de oxígeno relativamente alto (similar al actual) aparecen potencialmente como los más afectados tanto por los efectos directos del flash como por el efecto persistente del UV solar.
3. Entre las principales afectaciones de un BRG sobre la biosfera se encuentran el daño al ADN y los procesos de inhibición a los productores primarios tanto en el fitoplancton marino y como para el caso de las plantas terrestres.

4. La opacidad asociada al NO$_2$ en la región visible podría potenciar el efecto dañino del incremento del UV solar al inhibir, en alguna medida, el papel de los mecanismos reparadores fundamentalmente el de foto-reactivación.

# LA COMPLEJIDAD

# DE LA VIDA

## 3. LA COMPLEJIDAD DE LA VIDA

En este capítulo se examinan aspectos referentes a la dinámica de los ecosistemas estresados por los excesos de RUV que pueden ser, consecuencia directa de un BRG o de la explosión de una supernova cercana. Adicionalmente, se introducen también elementos sobre el origen e importancia de los ciclos biológicos en el contexto de las teorías sobre el surgimiento de la vida, resaltando las complejidades inherentes en este tipo de estudios.

### 3.1 Brotes de rayos gamma a nivel de los ecosistemas

En los capítulos anteriores hemos restringido nuestra atención, fundamentalmente, a la acción de un exceso de radiación ultravioleta sobre los llamados productores primarios. Sin embargo, resulta prácticamente imposible establecer una modelación más detallada de este fenómeno sin reconocer explícitamente la estructuración de la biosfera en ecosistemas y su continua interrelación con el clima. Muchos son los factores que conspiran contra un estudio de esta naturaleza destacándose, la elevada variabilidad en la sensibilidad a la radiación reportada para diferentes especies, pudiendo variar incluso, dentro de los organismos de una misma especie, el nivel de conocimiento muchas veces limitado que se dispone y los posibles umbrales y efectos no lineales asociados a este tipo de sistemas (Scheffer colaboradores, 2001; Van nes y Scheffer, 2004). Si bien y como una primera aproximación, pudiera ser aceptable considerar que las principales alteraciones ocurran inicialmente sobre los productores primarios de las biosfera (fitoplancton, algas, plantas superiores), al constituir estos la base de la cadena alimenticia, cualquier perturbación en ellos debe reflejarse de una manera más bien complicada en los niveles tróficos superiores (herbívoros, carnívoros, omnívoros). Es importante aclarar, que los excesos de radiación ultravioleta pueden afectar directamente a organismos expuestos a la radiación pertenecientes a otros niveles tróficos, fundamentalmente a aquellas en estado larvario o adulto (Vincent y Neale, 2000), ocasionando directamente la muerte o enfermedades tan extendidas como el eritema o el cáncer de piel. Pese a ser un problema científico bien establecido, la amplia diversidad de ecosistemas y las complejidades intrínsecas de este problema hacen que las posibles respuestas ante un exceso de radiación ultravioleta sea considerada, hasta hoy, una cuestión abierta (Hader y colaboradores, 2007).

En otro ámbito, como la biosfera regula en buena medida los niveles atmosféricos de importante gases tipo invernadero como son el $CH_4$, el $CO_2$, y los propios niveles de $O_2$ en escalas geológicas, las perturbaciones por exceso de radiación ultravioleta tanto de origen solar o extrasolar, pueden potencialmente llevar a cambios climáticos globales (Thomas y colaboradores, 2005) con las consecuencias ecológicas correspondientes. De hecho, reconocer la compleja interacción biosfera clima se convierte en un elemento clave para comprender con claridad, la evolución biogeoquímica de un planeta como la Tierra.

Para estudiar los efectos de la incidencia de un BRG a nivel regional o local, es importante dada su amplia diversidad, el modelado del exceso de radiación ultravioleta en varios ecosistemas. En este trabajo en particular se han elegido los lagos. Una de las razones para esta elección es que el modelo seleccionado de lago describe con éxito el proceso de eutrofización (exceso de enriquecimiento por los nutrientes, principalmente nitrógeno y fósforo) y se ha previsto, que uno de los efectos atmosféricos de un BRG sería una lluvia de compuestos de nitrógeno, lo que contribuiría a la eutrofización de los ecosistemas terrestres (Thorsett, 1995; Thomas y colaboradores, 2005). Esperamos entonces que nuestros resultados puedan resultar una medida aproximada de lo que podría ocurrir en una considerable fracción de las aguas continentales y los ecosistemas costeros después de la incidencia de la perturbación debida al BRG, ya que muchos de estos sistemas, a menudo, muestran ya un cierto grado de eutrofización debido fundamentalmente a los arrastres de materia orgánica y suelo de las tierras aledañas.

Por otro lado, pese a su carácter marcadamente local, un proceso como el de la eutrofización muestra una dinámica considerablemente compleja donde son posibles estados alternativos (bi-estabilidad), oscilaciones, histéresis, etc., comunes en otros ecosistemas de mayor relevancia global. Estas características los sitúan como un modelo adecuado para sondear desde un punto de vista cualitativo, posibles respuestas de otros ecosistemas bajo perturbaciones similares.

Admitimos que un modelado más exacto de la acción de un exceso de radiación ultravioleta a nivel de los ecosistemas requiere modelos específicos y de un elevado nivel de información para cada sistema en cuestión, tanto para el caso de los ecosistemas acuáticos como terrestres, tarea que se propone complementar en trabajos futuros.

### 3.1.1 El modelo integral de simulación acuática

El Modelo Integral de Simulación Acuática (CASM en la literatura) ha descrito con éxito las características principales del proceso de eutrofización en lagos reales (Amemiya y colaboradores, 2007). Este proceso está asociado, como dijimos, al sobre-enriquecimiento por nutrientes, principalmente fósforo y nitrógeno, con el consiguiente aumento de los niveles de fitoplancton, mientras que otras especies, tales como peces y zooplancton pueden escasear considerablemente bajo estas condiciones. Como dijimos en el punto anterior, la eutrofización por la deposición de nitratos es una de las posibles consecuencias de un BRG incidente en nuestra atmósfera, haciéndolo un proceso atractivo para nuestros propósitos. En este modelo hay una entrada externa del nutriente limitante para el ecosistema $I_N$, el cual en nuestro caso incluiría la deposición atmosférica de nitratos por lluvia posterior al evento. La ecuación (3.1) a continuación representa la dinámica de nutrientes en el ecosistema, donde $r_N$ es la tasa de pérdida de nitrógeno por diversas causas (por ejemplo, sedimentación, flujo, etc.), mientras que el tercer término de la derecha modela el consumo de nutrientes por parte de los consumidores primarios (fitoplancton $X$). La forma de este término se inspira en la cinética de Michaelis-Menten, comúnmente aplicada para modelar de manera sencilla algunos procesos enzimáticos. En nuestro caso, $\gamma$ es el cociente de la masa de nutrientes (masa de nitrato) a la biomasa, $r_1$ es la máxima tasa de crecimiento del fitoplancton y $k_1$ una constante de media-saturación (cuando $N = k_1$, todo el término se dividirá por dos después de cancelar $N$, de ahí la denominación de "media-saturación"). Por último, el cuarto término del lado derecho de la ecuación (3.1) representa la entrada de nutrientes $N$, a través de la descomposición de detritos $D$, teniendo en cuenta que $d_4$ es la tasa de descomposición de $D$. Llegamos así a

$$\frac{dN}{dt} = I_N - r_N - \frac{\gamma r_1 NX}{k_1 + N} + \gamma d_4 D \qquad (3.1)$$

La producción primaria del ecosistema está representada por la ecuación (3.2) a continuación, donde el fitoplancton $X$ consume nutrientes a través del primer término del lado derecho (se compárese con el tercer término del lado derecho de (3.1)), y el segundo término muestra cómo el zooplancton $Y$ se alimenta del fitoplancton. En este término, $f_1$ es la tasa de alimentación de zooplancton y $k_2$ es la constante de media-saturación para este término (porque cuando $X^2 = k_2$, la cancelación de $X^2$ asegura que todo el término se divide por dos). El último término del lado derecho de la ecuación contiene la mortalidad del fitoplancton $d_1$ y su velocidad de eliminación del ecosistema $e_1$. Tenemos así

$$\frac{dX}{dt} = \frac{r_1 NX}{k_1 + N} - \frac{f_1 X^2 Y}{k_2 + X^2} - (d_1 + e_1)X \qquad (3.2)$$

La ecuación (3.3) a continuación representa la dinámica del consumidor primario, el zooplancton $Y$. El primer término del lado derecho muestra cómo se alimenta del fitoplancton (compárese con el segundo término del lado derecho de la ecuación anterior), mientras que el tercer término dice cómo el zooplancton es depredado por el consumidor secundario, los peces zoo-planctívoros $Z$. El parámetro $\eta$ representa la eficiencia de la asimilación del zooplancton, y el sentido de los otros parámetros se puede deducir fácilmente de las explicaciones dadas para las dos primeras ecuaciones. Escribimos que

$$\frac{dY}{dt} = \frac{\eta f_1 X^2 Y}{k_2 + X^2} - \frac{f_1 Y^2 Z}{k_3 + Y^2} - (d_2 + e_2)Y \qquad (3.3)$$

La dinámica de los consumidores secundarios, los peces zoo-planctívoros, se indica en la próxima ecuación. Aquí la introducción de un nuevo parámetro $Z^*$, la biomasa inferior de equilibrio de los peces zoo-planctívoros, evita la situación irreal de versiones anteriores del CASM, en los que los peces pueden aparecer de los estados en los que ya estaban extinguidos.

$$\frac{dZ}{dt} = \frac{\eta f_2 Y^2 Z}{k + Y^2} - (d_3 + e_3)(Z - Z^*) \qquad (3.4)$$

Por último, debemos considerar que hay fuentes de detritos $D$ en el ecosistema (materia fecal y cadáveres $X$, $Y$ y $Z$), cuya descomposición devuelve los nutrientes al ecosistema. Esto es muy importante en todos los ecosistemas: una importante fracción de los nutrientes se devuelve al ecosistema a través de la descomposición de las heces y los seres muertos, como se indica en la ecuación 3.5

$$\frac{dD}{dt} = \frac{(1-\eta)f_1 X^2 Y}{k_2 + X^2} + \frac{(1-\eta)f_2 Y^2 Z}{k_3 + Y^2} + d_1 X + d_2 Y + d_3 Z - (d_4 + e_4)D \qquad (3.5)$$

Recordamos que los $d_i$ son las tasas de mortalidad o descomposición y $e_i$ son las tasas de remoción del sistema. Como puede verse de las ecuaciones (3.1) - (3.5), CASM posee cinco variables dinámicas y 19 parámetros (consultar Tabla 3.1 en el Anexo A para la parametrización empleada). En general, referimos al lector interesado al trabajo de Amemiya y colaboradores (2007) para más detalles.

### 3.1.2 La inclusión del transporte radiativo en el Modelo Integral de Simulación Acuática

La formulación del modelo CASM anterior no tiene en cuenta la distribución vertical de los seres vivos en la columna de agua. Esta es una omisión importante a la hora de considerar cualquier situación de estrés de RUV, dada la atenuación de la radiación debida a la absorción y dispersión en la columna de agua. Para tener en cuenta esto consideramos al fitoplancton como el único nivel trófico necesariamente bajo tensión por la radiación UV, ya que está obligado a tener una adecuada exposición solar con el fin de realizar la fotosíntesis (Neale y colaboradores, 2003). Podemos entonces imaginar todo el fitoplancton a una cierta profundidad efectiva estará expuesta al aumento de los niveles de RUV después de un BRG. Así, para incluir el papel de algunos componentes del ecosistema como cobertores del UV en la columna de agua (detritos y fitoplancton en sí mismos), hemos modificado el modelo CASM haciendo que el coeficiente de la tasa de mortalidad del fitoplancton ($d_i$) ya no una constante, sino una función explícita de esos componentes de la forma

$$d_i = d \times e^{-h_X X - h_D D} \qquad (3.6)$$

La dependencia exponencial anterior es motivada por la conocida ley de Beer para la absorción de la luz por soluciones líquidas $h_X$ y $h_D$ son los coeficientes de atenuación de la radiación UV por fitoplancton y detritos, mientras que $d$ es un coeficiente de tasa de letalidad del fitoplancton cuando no se considera el bloqueo de rayos UV.

La forma propuesta para modelar la atenuación del UV (ecuación 3.6) está en consonancia con los resultados de una larga lista de estudios experimentales realizados con este fin (ver para este y otros efectos, Lange y colaboradores, 2003). Hoy en día, se encuentra bien establecido el papel como agentes bloqueadores de la RUV que tienen pequeñas concentraciones de compuestos orgánicos disueltos (COD) o particulados (COP). Los niveles de estos compuestos pueden influir directamente en las distribuciones no solo del fitoplancton sino de organismos superiores en la cadena trófica como el zooplancton y los peces. Es importante destacar, que en el caso de los organismos superiores son comunes las estrategias de migración vertical con vistas a disminuir los daños asociados al UV (Hylander y Hansson, 2010; Leech y colaboradores, 2005). Esta situación suele ser mucho más notable en el caso de los lagos cristalinos debido a la mayor penetrabilidad del UV en la columna de agua.

### 3.1.3 **Principales resultados para escala regional**.

Para tener en cuenta los efectos combinados de la disminución de la capa de ozono y la reducción espectral de la luz solar, nuestra modificación del modelo CASM para los lagos se exploró con incrementos del coeficiente de la tasa de mortalidad del fitoplancton ($d_i$) sin considerar aún los efectos de fotoprotección explícitamente. Los resultados de esta exploración se expresan mediante un diagrama de bifurcaciones.

En la Figura 3.1 se muestra cómo el comportamiento cualitativo del modelo cambia en función del parámetro $d_i$ donde las líneas sólidas representan estados estables u oscilatorios mientras las de trazos representan estados transitorios. De su análisis se desprende que cuando el parámetro $d_i$ aumenta hasta 0,105, sólo un 5% superior al valor de referencia de 0,1 utilizado por Amemiya y colaboradores (2007), el estado estacionario surge claramente como oscilatorio en una llamada bifurcación de Hopf. Para valores más altos (alrededor de 0,125), la bi-estabilidad del sistema se rompe y el estado oscilatorio surge como la única posibilidad. Tales estados alternativos son exhibidos por el CASM para otras regiones de los parámetros (Amemiya y colaboradores, 2007).

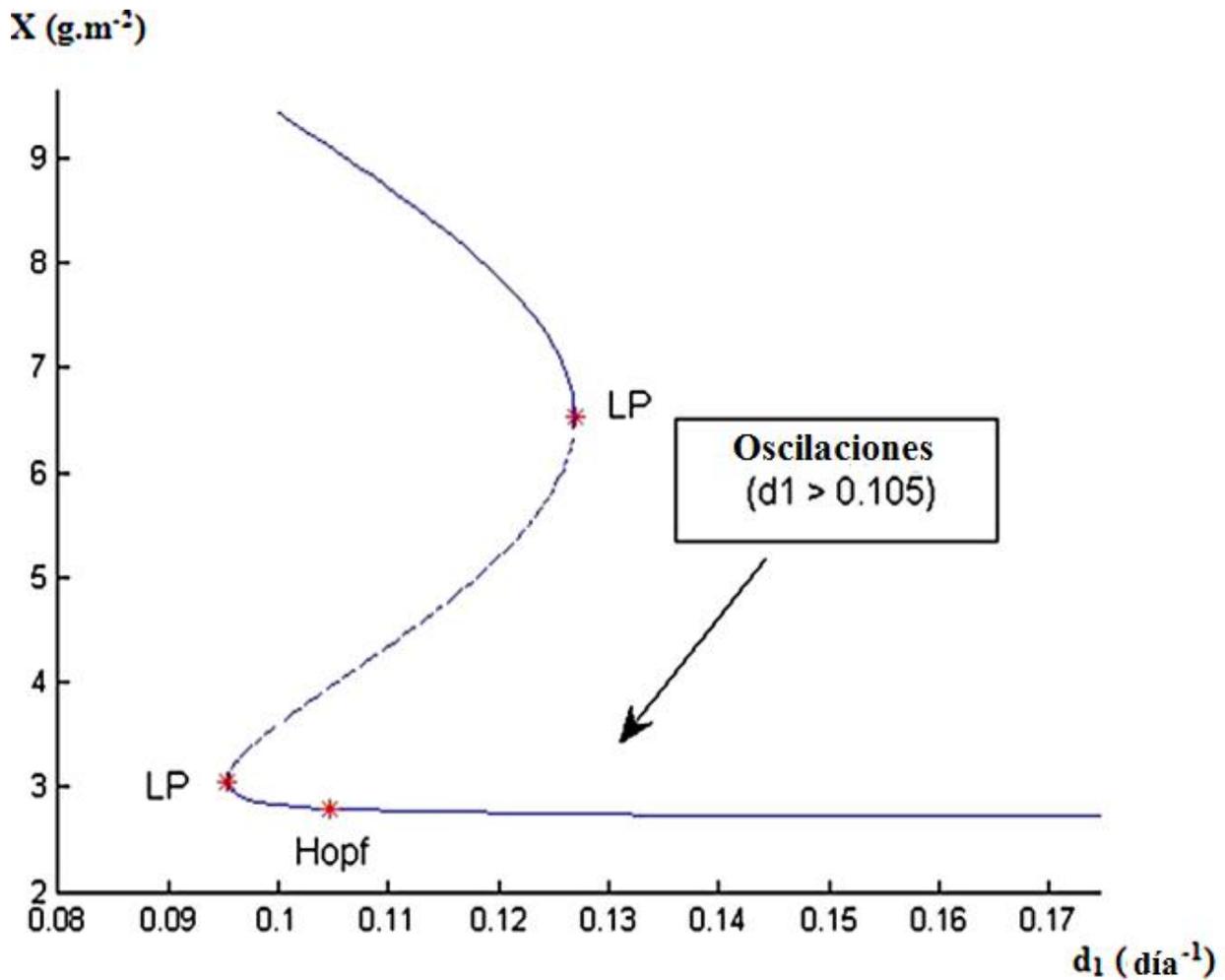

Fig. 3.1 Diagrama de bifurcaciones para el modelo CASM en función del parámetro de tasa de mortalidad del fitoplancton ($d_1$).

Un análisis del transporte radiativo en los regímenes de oscilación parece interesante debido a que las propiedades ópticas de la columna de agua están continuamente variando en el tiempo. Algunos de los componentes materiales como el detritos ($D$) y el fitoplancton ($X$) tienen el papel de protección adicional contra el UV para las principales especies subacuáticas. Teniendo en cuenta nuestra expresión modificada para el coeficiente de la tasa mortalidad (ecuación 3.6) y considerando contribuciones iguales para la atenuación de los fotones UV por el fitoplancton y detritos ($h = h_X = h_D$), encontramos el comportamiento mostrado en la Figura 3.2.

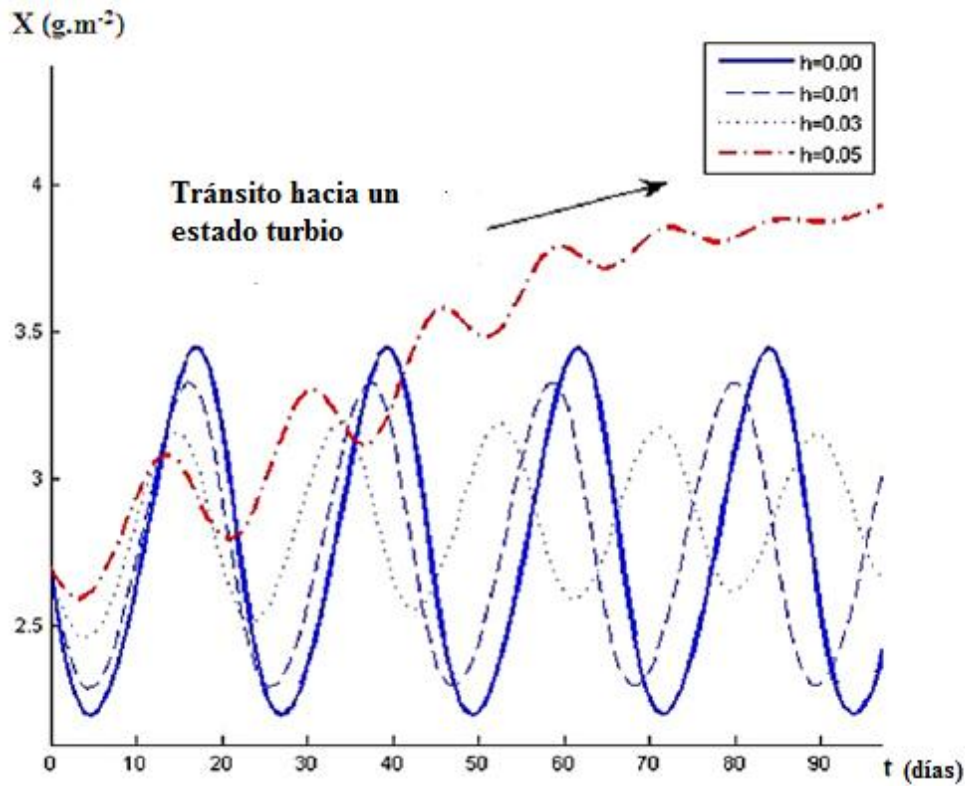

Fig. 3.2 Efectos de la auto-protección al UV sobre la dinámica del estado cristalino.

Ahora, de acuerdo con la figura 3.2, si la auto-protección no es demasiado alta, el régimen oscilatorio alrededor del estado cristalino persiste, con pequeñas correcciones en la amplitud y período de oscilación. Si la auto-protección alcanza cierto valor umbral, la población de fitoplancton en el tiempo sufre una recuperación progresiva regresando el ecosistema al régimen turbio original.

Los resultados anteriores ilustran la complejidad del proceso de predicción de cómo los ecosistemas terrestres se recuperarían si fuesen alcanzados por un BRG. El estado hacia el que un lago dado evolucionará podría depender de varias variables y parámetros, donde el fitoplancton parece jugar un papel preponderante. Sin embargo, un modelado más exacto de la recuperación de los ecosistemas acuáticos después de un BRG necesita un estudio más detallado del comportamiento de las especies más comunes de fitoplancton bajo estrés RUV, y también de la inclusión de otras variables ambientales en consideración.

Por otro lado, transiciones abruptas o cambios de regímenes entre posibles estados alternativos a escala global pueden generar un conjunto de serias implicaciones ecológicas a escala planetaria. La posibilidad de un cambio irreversible en la biosfera a nivel global podría tener consecuencias catastróficas para muchos ecosistemas terrestres incluyendo a la especie humana y debe ser un aspecto fundamental en la mayor parte de las políticas y estrategias de protección y recuperación de ecosistemas y del medio ambiente en general, que en una buena parte de ellas, presuponen un carácter lineal (Van Nes y Scheffer, 2004).

## 3.2 **Complejidad a nivel celular**

La riqueza en el comportamiento exhibido por los sistemas biológicos puede considerarse, en última instancia, como una expresión de la naturaleza compleja de los seres vivos. Aunque en el primer capítulo se introdujo el tema, básicamente abordamos condiciones generales para el surgimiento de la vida en un contexto astrofísico o cosmológico, sin dedicar mayor atención al fenómeno de la vida y a las principales ideas acerca de su surgimiento. La inclusión de estos tópicos, de trascendencia universal, nos permite ganar en completitud de nuestro trabajo teniendo en cuenta que consideramos escenarios terrestres en el periodo Arcaico o Hadeico hace unos 3.5 o 4.0 Ga atrás.

Particularmente en el ámbito terrestre, pese al éxito posterior de los organismos fotosintéticos, existe cierto consenso en que las primeras formas de vida empleaban la quimio-síntesis como vía primaria para obtener la energía. Anterior a estos periodos, comienza lo que se ha dado a conocer la era prebiótica, marcada por la ausencia de cualquier forma o expresión de vida, independientemente del grado de organización.

Grande es el nivel de incertidumbre con respecto a esta etapa, no solo en lo referentes a aspectos climáticos no resueltos aún, como la conocida paradoja del sol pálido (aspectos del tema en Grenfell y colaboradores, 2010), sino también a los procesos físico-químicos y de auto-organización que dieron lugar a la emergencia de un fenómeno, tan complejo y tan poco entendido, como es la vida. Varios son los enfoques recogidos en la literatura sobre esta problemática que van desde los más clásicos, como la formación de compuestos orgánicos claves en línea con el famoso experimento de (Miller y Urey, 1953), hasta los que implican principios aún no establecidos que involucran áreas fundamentales de la mecánica cuántica, los sistemas complejos y la teoría de la información (para una discusión general sobre estas temáticas consultar Davies, 2004).

Durante esta etapa, la radiación juega un papel decisivo en la mayor parte de los procesos de síntesis y degradación de compuesto orgánicos que se establecen fundamentalmente en la atmosfera y en el océano. Es notable el número de mecanismos de reacción propuestos en la literatura donde la interacción con la radiación, básicamente con el UV, aparece explícitamente en alguna etapa (Cleaves y Miller, 1998). En algunos casos, la producción de estos compuestos podría haber alcanzado niveles notables llegando a crear una densa niebla capaz de cubrir el planeta (Pavlov y colaboradores, 2001).

Adicionalmente, los efectos directos de la radiación podrían estar conectados con el origen de la homoquiralidad (Fitz y colaboradores, 2007), propiedad distintiva de moléculas biológicas consideradas claves como los azúcares y aminoácidos. La homoquiralidad se basa en la propiedad de especies moleculares idénticas, no superponibles (enantiomeros), de girar el plano de polarización de la luz en un sentido bien determinado, siendo los azúcares biológicamente activos únicamente derechos y los aminoácidos izquierdos. Hechos astronómicos como la identificación de una intensa fuente de luz circularmente polarizada en una de las nubes de polvo de la constelación Orión (Baley y colaboradores, 1998), similar a la que dio origen a nuestro sistema solar, sustentan este tipo de conjeturas.

Sin embargo, probablemente el efecto más notorio de la radiación es su ya mencionado carácter cíclico en días y noches. Ciclos que se traducen en la mayor parte de las reacciones atmosféricas que comienzan en procesos fotoquímicos y en los ya mencionados ritmos circadianos que tienen lugar en los organismos vivos.

### 3.2.1 Oscilaciones químicas y bioquímicas. Origen y Clasificación

Hoy día, la presencia de oscilaciones químicas o bioquímicas puede ser entendida al menos de dos formas, como un proceso exógeno o endógeno. Los primeros son propiciados por agentes externos que pueden incluir fluctuaciones de parámetros ambientales tales como la iluminación, temperatura, pH u otros sobre un sistema de reacciones químicas determinado, no necesariamente de naturaleza biológica. Entre estos, las contribuciones más importantes parecen estrechamente relacionadas con la extensión del ciclo diario de luz y oscuridad (de 15 h al inicio del arcaico y 24 h ahora). Tal influencia, de efectos notables en la dinámica del ozono y de otras muchas especies en la química atmosférica, parece decisiva en la expansión de los ritmos circadianos en plantas, animales y microorganismos como cianobacterias (Para más elementos de los ritmos circadianos ver Mihalcescu y colaboradores, 2004; Lakin-Thomas y Brody, 2004; Liebermeister, 2005). Su origen, conjuntamente con el de las proteínas fotosensibles, aparece tempranamente en las células primitivas con el propósito de proteger la replicación de DNA de altas dosis de radiación ultravioleta durante el día en el eón Hadeico o Arcaico. En consecuencia con tales criterios, el ritmo emerge como un factor clave en la regulación de procesos bioquímicos dentro de un individuo, así como en la coordinación ante las fluctuaciones medio ambientales. Otras fuentes de ciclicidad a escalas de tiempo considerablemente mayores, podrían tener alguna repercusión directa o indirecta en el origen, distribución y evolución de la vida como pueden ser los ciclos solares, los llamados ciclos de Milankovich o el tránsito del sistemas solar por lo brazos espirales de nuestra galaxia.

A diferencia de las anteriores, las llamadas oscilaciones endógenas se originan como el resultado de la compleja red de interacciones (metabólica en el caso de los seres vivos) con múltiples rutas y lazos de retroalimentación entre los diferentes componentes (metabólicos, enzimas, etc) y estructuras (ver Micheva y Roussel, 2007).

Aunque inicialmente introducido en la química con la reacción de Belousov-Zhabotinsky (BZ) para describir el curso de la oxidación de los iones $Ce^{3+}$ y $BO_3^-$ en solución ácida, el término reacciones oscilantes rápidamente alcanzó sus mejores aplicaciones en los sistemas bioquímicos y biológicos.

Las oscilaciones, tanto de carácter exógeno como endógeno, aparecen como procesos genéricos (McKane y colaboradores, 2007) en los sistemas biológicos, ocurriendo las mayoría de la veces de manera combinada, o sea, por ejemplo una oscilación endógena modulada por una fuente exógena. Quizás de los mejores ejemplos de este comportamiento sea la fluctuación periódica exhibida por ATP y glucosa durante el ciclo de glucólisis. De manera general, las oscilaciones se consideran como un componente crucial de importantes procesos celulares que no solo comprenden las rutas metabólicas, sino también la reproducción e inclusive la señalización (Igoshin y colaboradores, 2004).

### 3.2.2 Estableciendo un modelo Vida.

Basados en las consideraciones citadas anteriormente podemos dar por supuesto que las oscilaciones son una propiedad ancestral de los sistemas vivos, y esta suposición nos permite (en forma más bien drástica) clasificar un sistema simple como vivo (oscilando) o muerto (no-oscilando). De hecho esta definición sabemos resulta demasiado simplificada para organismos modernos, pero creemos que podría ser de ayuda para estudiar y clasificar algunos sistemas simples, probablemente parecidos a la primeras expresiones " abióticas " de la vida.

Elegir un esquema como este evita, en buena medida, los problemas relacionados con establecer una definición universalmente aceptada del concepto de vida (ver Joyce y Orgel, 1993 ; Lazcano, 2010) acercándose a una línea de pensamiento introducida inicialmente por (Kauffman, 1995) y a escenarios del tipo "El metabolismo primero". Conjuntamente con el fenómeno de las oscilaciones, la capacidad de sincronización (Bier y colaboradores, 2000), la homoquiralidad (Plankensteiner y colaboradores, 2004) y la emergencia de complejas redes de interacción (Fell y Wagner, 2000) aparecen como características distintivas de todos los seres vivos, sugiriendo una conexión mucho más intrínseca y profunda con el fenómeno de la vida en sí mismo.

Otro elemento clave que consideremos en nuestro modelo es la existencia de una membrana o pared celular. En las células modernas dicha membrana es indispensable para el funcionamiento celular, controlando el flujo de agua, el intercambio de sustancias y delimitando las dimensiones efectivas de la célula. Entre los roles de la membrana se encuentra además, el resguardo de las estructuras internas de la célula de la influencia nociva del medio ambiente (para más detalles Zonia y Munnik, 2007).

No está completamente claro si el antepasado de la célula primitiva tuvo una pared celular o no, pero parece un hecho probable teniendo en cuenta las múltiples funciones que esta asume en los organismos modernos. Varios compuestos han sido propuestos como posibles candidatos para conformar la membrana en las células primitivas (Ourisson y Nakatani 1994; Deamer y colaboradores, 2002 para ver detalles). Desafortunadamente no hay una alternativa clara hasta ahora para este cuadro, y la estabilidad de las estructuras formadas se sabe son extremadamente dependientes de parámetros ambientales como el pH del medio. También se ha alegado en los últimos años, que algunos de los más importantes compuestos para formar estas estructuras pudieron tener un carácter exógeno, condicionados por la transferencia de material orgánico proveniente del espacio exterior (Bernstein y colaboradores, 2001, 2002; Muñoz Caro y colaboradores. 2002; Dworkin y colaboradores 2001).

Básicamente nuestro modelo rudimentario de célula primitiva estaría conformado por un sistema de reacciones químicos capaz de oscilar, limitado por una membrana semipermeable y flexible que permite el transporte pasivo de determinadas sustancias. El volumen de la célula viva, a semejanza de las células reales, será de magnitud variable en correspondencia con los gradientes específicos (presión osmótica) a través de la membrana afectando la dinámica del sistema original (ver Zonia y Munnik, 2007). En este sentido, las propiedades intrínsecas de la membrana determinarán la factibilidad del estado oscilante (célula viva).

Aunque en principio es posible implementar las ideas discutidas con anterioridad en cualquier oscilador hipotético o real, por su importancia consideraremos básicamente el llamado oscilador glucolítico.

### 3.2.3 La glucólisis como un caso particular

Consideremos una modificación de la versión mínima del modelo propuesto originalmente por (Bier y colaboradores, 2000) con la meta de modelar la sincronización oscilatoria del proceso glucolítico en células de levadura (ecuaciones 3.7, 3.8)

$$\frac{dG}{dt} = V_{in} - k_1 GT - G\frac{dV}{dt} \qquad (3.7)$$

$$\frac{dT}{dt} = 2k_1 GT - \frac{k_p T}{K_m + T} - T\frac{dV}{dt} \qquad (3.8)$$

El último término en cada ecuación se incluye con el objetivo de considerar explícitamente el efecto de un volumen variable sobre la dinámica en correspondencia con los trabajos previos de (Pawlowski y Zielenkiewicz, 2004). La forma específica del término dV/dt puede ser estimada a partir de la velocidad con que cambia la concentración total de la especies oscilantes dN/dt, donde el valor de N será la suma de las diferentes concentraciones en el sistema anterior, en este caso N=G+T. En correspondencia con esto, considerando el sistema anterior (ecuaciones 3.14-3.15) y después de algunos pasos algebraicos, es posible encontrar una expresión para el término como

$$dV/dt = \frac{r}{1+r(T+G)}\left(V_{in} + k_1 GT - \frac{k_p T}{K_m + T}\right) \qquad (3.9)$$

En esta expresión, el parámetro r tiene la función de codificar el comportamiento específico de la membrana frente a cambios de volumen o a procesos difusivos a través de ella. Si superponemos que su valor es pequeño, es posible, considerando un desarrollo en series de potencias encontrar una expresión simplificada para el término de cambio de volumen de la forma

$$dV/dt \approx r\left(V_{in} + k_1 GT - \frac{k_p T}{K_m + T}\right) \qquad (3.10)$$

A partir de razonamientos similares, es posible realizar estimaciones de otros efectos como la magnitud y el sentido de los flujos de solvente en función del comportamiento de la presión osmótica, teniendo en cuenta que para soluciones diluidas se cumple que

$$J_{solv} \sim P_{osm} \approx \Delta C_N \cdot R \cdot T \qquad (3.11)$$

dónde $P_{osm}$ es la presión osmótica. $\Delta C_N$ es la diferencia de concentración dentro y fuera la célula de las especies oscilantes, $R = 8.314$ J K$^{-1}$ mol$^{-1}$ es la constante universal del gas y T es la temperatura expresada en grados Kelvin (K). Varios son los análisis que se pueden hacer a partir de la ecuación entre los que se destacan la existencia de un flujo límite cuando la presión osmótica sobrepase determinado umbral o cuando la frecuencia de las oscilaciones sea demasiado elevada para que la condición de equilibrio referida en la ecuación anterior llegue a ser aplicable.

Básicamente, con la introducción del nuevo parámetro, nuestro objetivo se limita ahora a explorar sus efectos sobre las oscilaciones originales del sistema. Para ello consideraremos primeramente el uso de la expresión aproximada (ecuación 3.9) y con posterioridad mediante el empleo de la expresión general (ecuación 3.10). La parametrización empleada en ambos casos se corresponde a la originalmente empleada por (Bier y colaboradores, 2000) consistente con un estado oscilante (consultar tabla 3.2 en el Anexo B), coincidente con la nueva versión para el valor del parámetro $r = 0$.

**Un análisis aproximado**

Exploremos entonces, desde el punto de vista numérico, el comportamiento del modelo aproximado (ecuación 3.9) para valores crecientes del parámetro r. Los resultados de este procedimiento se muestran en la figura 3.3.

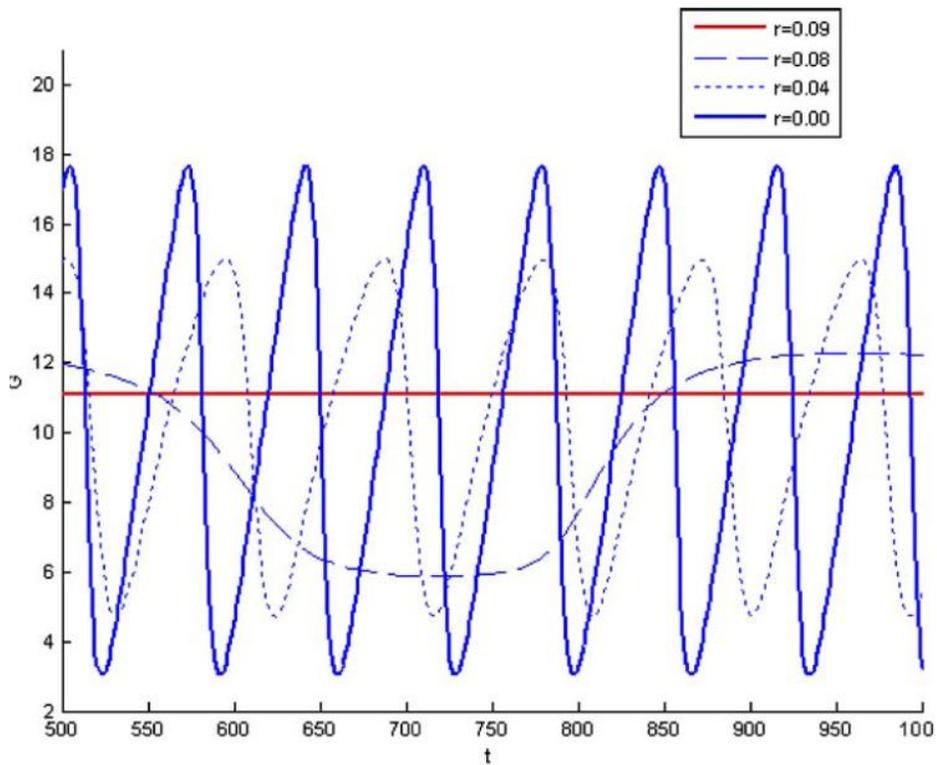

Fig. 3.3 Patrones de oscilación para valores crecientes del parámetro r en el caso del oscilador glucolítico. La línea roja, paralela al eje temporal, implica la existencia de un estado muerto (no-oscilante)

En correspondencia con el comportamiento exhibido en el gráfico, el modelo aproximado sugiere que las membranas con baja rigidez o con poca resistencia al paso del solvente, limitan la viabilidad de los estados oscilantes o vivos según nuestro criterio. En este caso se puede observar como el periodo crece, mientras la amplitud de las oscilaciones decrece ante incrementos pequeños del parámetro. Nótese además, que por encima de determinado umbral el sistema pierde completamente sus características oscilantes pasando a lo que llamaríamos un estado muerto.

**El Caso general**

Pese a que los resultados obtenidos para el caso anterior son razonables, recordemos que su validez puede encontrarse limitada a las regiones donde el parámetro r pueda considerarse realmente pequeño. Si analizamos las implicaciones sobre la dinámica de la expresión general (ecuación 3.10) obtenemos un comportamiento algo diferente. Nuevamente los incrementos del parámetro r conspiran contra el carácter oscilatorio del sistema que se traducen en un incremento paulatino del periodo de las oscilaciones (ver figura 3.4). Sin embargo, a diferencia del caso aproximado, no llegan a ocurrir cambios cualitativos de las soluciones (bifurcaciones) para incrementos notables del parámetro. Este comportamiento implica, al menos en este tipo de construcción, la robustez del fenómeno de las oscilaciones frente a los procesos de cambio de volumen o del transporte pasivo a través de la membrana.

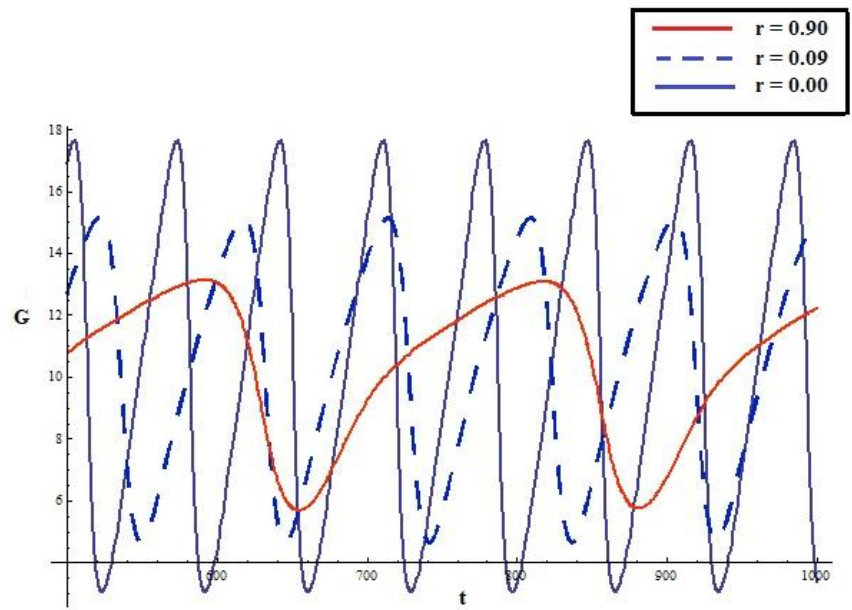

Fig. 3.4 Incremento paulatino del periodo de las oscilaciones con el incremento del parámetro r

Sin embargo, si tenemos en cuenta el presumible rol que desempeñan las oscilaciones en la dinámica celular y consideramos además la vida como un fenómeno limitado en el tiempo, entonces un incremento considerable del período de las oscilaciones puede considerarse, en un sentido práctico, como un estado no oscilante, o sea, muerto en nuestro criterio.

Si uno extiende este estudio para incluir otros osciladores químicos o bioquímicos (para el caso de la mitosis consultar Goldbeter, 1991 ) bien establecidos y hace un análisis algo más cuidadoso mediante el estudio numérico (Dhooge y colaboradores, 2006; Doedel, 1986) de posibles bifurcaciones asociadas al parámetro r, descubre una tendencia general a la pérdida del comportamiento oscilatorio. Sin embargo, los efectos sobre los patrones de la oscilación si pueden variar sustancialmente de un oscilador a otro sin mostrar una regla o tendencia bien establecida (Martín y colaboradores, 2009).

Las motivaciones en considerar la glucolisis como caso distintivo se debe fundamentalmente a su presencia en casi todos los organismos como una fuente principal de energía química y ser, quizás, la más antiguas vía metabólica (Fell y Wagner, 2000). Estas características la convierten en un proceso particularmente interesante del punto de vista del astrobiológico. El proceso en sí mismo consiste en la ruptura gradual de las reservas de glucosa para liberar energía en forma de moléculas ATP. Está bien documentado el hecho que durante el proceso, las concentraciones de los diferentes metabolitos (ATP, glucosa) exhiben oscilaciones respecto a cierto valor de concentración medio. Se conoce además que dichas oscilaciones poseen la habilidad de sincronizarse entre células diferentes. Para este trabajo, hemos elegido uno de los varios modelos que se han propuesto para estudiar y reproducir los patrones oscilatorios y los posibles mecanismos de sincronización.

Para terminar es bueno considerar dos elementos adicionales. Primero debe quedar claro que en nuestro análisis describimos sólo membranas pasivas, en el sentido que no están directamente acopladas al propio metabolismo de la célula. Obviamente, éste no es el caso de organismos modernos donde las estructuras de membrana se regeneran y crecen continuamente, pero aceptable en los comienzos, cuando su papel principal pudo estar limitado a actuar como simples receptáculos para los procesos "bioquímicos" incipientes.

El segundo elemento se relaciona con algunos aspectos termodinámicos de la vida y nuestros criterios específicos de vida y muerte. Desde un punto de vista estrictamente termodinámico, ambos estados estarían vivos en el sentido que los dos se encuentran fuera del equilibrio, existiendo flujos netos de masa y energía. Si nuestro criterio fuera acertado, entonces los sistemas vivos formarían una clase más bien limitada dentro de otra, mucho más general, que incluiría a todos aquellos sistemas alejados del equilibrio.

3.3 **Conclusiones**

Los resultados obtenidos en este capítulo muestran como la modelación de los sistemas biológicos es una tarea altamente compleja, no solo por el elevado número de variables involucradas o la carencia de datos sino por el carácter marcadamente no lineal de su comportamiento. Particularmente para el caso regional, hemos explorado los posibles impactos que provocan el efecto combinado de los excesos de radiación UV y el sobre-enriquecimiento de nutrientes vinculados a eventos como las BRG o las supernovas cercanas. En correspondencia con estos resultados, los efectos al parecer importantes sobre los ecosistemas estudiados, podrían exhibir un comportamiento más bien variado, dependiendo entre otros factores de la resistencia de las especies presentes (fundamentalmente del fitoplancton) y de la posibilidad de efectos de apantallamiento al exceso de UV en la columna de agua que reduzcan de manera considerable los niveles de exposición. Por otro lado hemos discutido el papel de los ciclos, importancia y modulación por efecto del transporte pasivo a través de las membranas. Las conclusiones fundamentales que se derivan de los análisis anteriores podrían resumirse como:

1. Las perturbaciones intensas en el régimen fotobiológico pueden alterar sensiblemente la dinámica de los ecosistemas llegando incluso a inducir cambios irreversibles en el estado de estos.

2. Las complejidades inherentes a los sistemas biológicos, tanto a nivel de sistemas (biosfera, ecosistema) como de organismos unicelulares, la carencia de datos fidedignos así como aspectos específicos del método de modelación ecológica, limitan en buena medida el alcance de este tipo de estudios.

3. Las propiedades de las membranas primitivas y el transporte pasivo pueden modular la emergencia de ciclos, elementos interesantes desde el punto de vista astrobiológico si tenemos en cuenta su importancia y extensión entre los organismos vivos.

# CONCLUSIONES

CONCLUSIONES

El papel de las radiaciones sobre el origen, instauración y evolución de la vida ha sido por varias décadas un tema activo de investigación, no solo en el marco de las investigaciones puramente astrobiológicas, sino en una buena parte de los trabajos astrofísicos y de las llamadas Ciencias de la Tierra. Enmarcado en este contexto, nuestro trabajo tiene la peculiaridad de abordar esta problemática desde una perspectiva general que involucra tanto los aspectos cosmológicos o galácticos del problema hasta aquellos de carácter planetario alcanzando incluso elementos de índole local o regional.

Desde el punto de vista más general posible, el fenómeno de la vida parece estar condicionado directamente con la dinámica y las leyes más generales que rigen nuestro universo y que han dado lugar a la formación de estructuras como nuestra galaxia. Es importante acotar que este carácter biofílico, expresado en otros conceptos como los de zona estelar o galáctica habitable, son consecuencia directa de la dinámica y no de un intervalo temporal específico. Tengamos en cuenta que las condiciones del universo temprano no eran en lo absoluto favorables para la vida. Un razonamiento similar es igualmente extensible tanto a la evolución de nuestra galaxia como a los orígenes del sistema solar.

Ya en un contexto planetario, el régimen fotobiológico emerge como una variable climática significativa en el comportamiento de la biosfera como sistema y particularmente en el caso de los productores primarios. Bajo las condiciones del régimen fotobiológico estacionario establecidas por una estrella como el Sol, las contribuciones relativas en las regiones UV y visible del espectro en superficie aparecen como elementos determinantes que condicionan, junto a otros factores tales como la circulación o presencia de bloqueadores del UV en el agua, la extensión, el hábitat de los organismos fotosintéticos así como la eficiencia del propio proceso de la fotosíntesis.

Por otro lado, las alteraciones bruscas del régimen fotobiológico asociados a eventos astrofísicos intensos como un BRG cercano puede ejercer un impacto significativo sobre varios componentes de la biosfera y particularmente sobre los productores primarios, variando en correspondencia de otros factores climáticos como los niveles de oxígeno y ozono libres en la atmósfera. Sin embargo, pese a ser un hecho factible, es muy difícil estimar de una manera clara, la posible repercusión que un evento de esta naturaleza sería capaz de desencadenar sobre la biosfera a nivel global, si tenemos en cuenta las complejidades intrínsecas manifiestas no solo a nivel de ecosistemas sino a nivel de la propia vida, interpretada como un fenómeno complejo. Los aspectos discutidos con anterioridad podrían resumirse a modo de conclusiones en:

- La dinámica del Universo en general y de la Vía Láctea en particular condicionan cierto nivel de biofilia que se manifiesta, de manera más notable, en aquellas regiones dentro de las llamadas zona estelar habitable y zona galáctica habitable.

- El régimen fotobiológico y sus posibles perturbaciones condicionan en buena medida la viabilidad, extensión y hábitat de los productores primarios fotosintéticos así como la eficiencia de procesos claves como es el caso de la fotosíntesis.

- Modelar de manera detallada el impacto que tienen eventos como los BRG sobre la biosfera constituye una tarea extremadamente difícil si tenemos en cuenta, además de las considerables incertidumbres climáticas aparejadas a estos fenómenos, el comportamiento altamente complejo que exhiben todos los sistemas biológicos y particularmente los ecosistemas.

# RECOMENDACIONES

RECOMENDACIONES

1. Extender los estudios sobre la influencia de las explosiones estelares en la tierra temprana con vistas a estimar posibles efectos fotoquímicos en atmósferas con composiciones variables de $CO_2$ y metano fundamentalmente.

2. Continuar el desarrollo de nuevos índices y criterios de daño biológico apropiados para este tipo de escenario.

3. Explorar otros ecosistemas modernos de interés bajo el efecto sostenido de un incremento de las radiaciones ultravioletas.

# REFERENCIAS BIBLIOGRÁFICAS

REFERENCIAS BIBLIOGRÁFICAS

# ANEXOS

GLOSARIO

**Principales términos de corte cosmológico y astrofísico empleados**

**Brotes de rayos gamma (BRG)**: Las mayores explosiones reportadas en el universo con fluencias del orden de unos $10^{44}$ J y una duración promedio del orden de algunos segundos a minutos. Se asocian a la interacción de sistemas binarios o al colapso de estrellas muy masivas. Se caracterizan además por que las emisiones ocurren fundamentalmente en la región gamma del espectro y son colimadas.

**Energía Oscura**: Componente mayoritario de naturaleza aún no establecida que representa prácticamente dos tercios del contenido de materia-energía del universo. Se considera el responsable de la aceleración expansiva de nuestro universo.

**Materia Ordinaria**: Denominación empleada para la materia de naturaleza bien establecida como son los bariones, fotones y neutrinos con vistas a diferenciarla de la llamada Materia Oscura.

**Materia Oscura**: Componente mayoritario de naturaleza poco establecida, de carácter atractivo y que representa la mayor parte de la materia no radiante que se agrupa en las galaxias, cúmulos y supercúmulos de galaxias. Sus efectos son ya notables en la dinámica galáctica.

**Modelo Cosmológico Estándar**: Criterio introducido originalmente por Einstein y ampliamente asumido por la cosmología moderna para describir las propiedades y evolución del universo a gran escala. En correspondencia con este, nuestro universo hoy en día puede considerarse plano, homogéneo e isótropo a escalas considerables.

**Principales términos de corte astrobiológicos**

**Borde rojo**: Comportamiento espectral típico exhibido por la atmósfera como consecuencia directa de la absorción de los pigmentos de la clorofila alrededor de los 0.7 μm (luz roja). Es considerado como una bioseñal superficial importante en los programas de búsqueda de vida extrasolar.

**Mundo ARN**: Forma parte de las teorías que intentan explicar el origen de la vida y pese a sus muchas limitaciones, se considera entre las alternativas más aceptadas por la comunidad científica. Desarrollada en sus orígenes por H.J. Muller (1926) supone que la información juega un papel preponderante y que inicialmente pudo estar codificada preferentemente en el ARN.

**Metabolismo Primero**: Forma parte de las teorías que intentan explicar el origen de la vida suponiendo que el metabolismo es un factor clave para este proceso. Sus fundamentos iniciales descansan sobre los trabajos de A. Oparin (1938) sobre los coacervados.

**Panspermia**: Teoría popularizada por Arrhenius (1903) que supone a la vida como un fenómeno no originario de nuestro Planeta, cuya semilla provino del espacio exterior con el impacto de asteroides o cometas.

**Zona estelar habitable**: Región alrededor de una estrella considerada favorable para la emergencia de vida introducida por J.F. Kasting. Entre los requerimientos más aceptados se encuentran los niveles de irradiancia adecuados sobre un planeta rocoso, capaz de retener atmósfera y agua líquida.

**Zona galáctica habitable**: Puede considerarse una extrapolación del concepto de zona galáctica habitable. Supone que las condiciones de vida en la galaxia varían significativamente siendo favorables en una zona intermedia del disco. Los criterios se basan fundamentalmente en la estabilidad (menor número de supernovas) y al nivel de metalicidad (entiéndase elementos diferentes al H y He).

**Términos de corte ecológico**

**Biosfera**: Términos de implicaciones planetarias usado para referirse a la interacción de todos los organismos vivos organizados en ecosistemas.

**Biota:** Denominación genérica usada para referirse a todos los organismos vivos sin especificaciones.

**Capa mezclada**: Se denomina a la región del océano con propiedades químico físicas homogéneas debido a los procesos de transporte de diferente naturaleza. Puede extenderse de escasos metros hasta profundidades considerables de cientos de metros.

**Detritus**: Nombre con que se designa la materia orgánica residual dentro de ecosistema determinado integrada fundamentalmente por las heces, organismos muertos, hojas y otras materias orgánicas en descomposición.

**Ecosistemas**: Relaciones que se establecen a escala regional o local entre las diversas especies y comunidades y con el medio ambiente. El cuidado y manejo de estos sistemas se considera un elemento indispensable en todas las políticas ambientales.

**Eón**: Intervalo de tiempo del orden de los Ga en que se divide la historia geológica de nuestro planeta para su estudio. Se reconocen cuatro eones que en orden cronológico ascendente serían Hadeico, Arcaico, Proterozoico y el Fanerozoico (actual)

**Eutrofización**: Proceso asociado al exceso de nutrientes fundamentalmente nitrógeno y fósforo y que se caracteriza por un incremento desmedido del fitoplancton. La eutrofización es la causa de la turbidez que exhiben muchos lagos y áreas costeras en la actualidad debido al uso excesivo de fertilizantes.

**Fitoplancton**: Integrado por un gran número de especies y organismos unicelulares entre los que se destacan diferentes tipos de cianobacterias y algas. Por ser los productores primarios más extendidos se consideran un elemento indispensable tanto dentro del ecosistema como a nivel climático global.

**Relaciones tróficas**: Dependencias de depredación (alimentación) que establecen diferentes especies de organismos, usualmente con grados de organización diferentes.

**Ritmos circadianos**: Periodicidad que manifiestan una amplia generalidad de procesos que ocurren en los organismos vivos motivadas fundamentalmente por la alternancia diaria de luz oscuridad entre otras variables climáticas. Es considerada como una estrategia adaptativa importante frente a los excesos de radiación UV.

**Términos de corte fotoquímico o fotobiológico**

**Factor de Amplificación Biológica (FAB):** Función de ponderación biológica que se construye con vistas a estimar los efectos del UV sobre determinado proceso orgánico. Incluye la acción de los mecanismos de reparación celulares.

**Factor de Amplificación de la Radiación (FAR)**: Función de ponderación biológica construida con vistas a estimar los daños por el incremento de UV asociados a las variaciones en el contenido de ozono. No incluye en su definición la acción de los mecanismos de reparación celulares.

**Fotoreactivación**: Es uno de los mecanismos más extendidos y eficientes de reparación de los daños al ADN por efecto del UV. El proceso depende de una enzima denominada fotoliasa y de luz en el rango de 350-450 nm (azulada)

**Funciones de Ponderación Biológica (FPB):** Funciones matemáticas construidas generalmente a partir de estudios experimentales para cuantificar los efectos de determinado agente (ejemplo UV) sobre determinada estructura o proceso orgánico como pueden ser el ADN o la fotosíntesis respectivamente.

**Radiación Fotosintéticamente Activa (RFA)**: Intervalo de longitudes de onda comprendido aproximadamente entre 400-700 nm que contribuye activamente al proceso de fotosíntesis.

**Zona Fótica**: Término empleado para referirse aquella región con niveles de luz suficiente como para realizar el proceso de fotosíntesis. Su amplitud depende directamente de aquellos factores que afectan la transparencia de la columna del agua.

**Apéndice A**

Parámetros empleados en el modelo CASM. Los valores fueron tomados de (Amemiya y colaboradores, 2007).

| Parámetro | Valor | Unidades |
|:---:|:---:|:---:|
| $I_N$ | 0.8 | $mgm^{-2}día^{-1}$ |
| $Z^*$ | 0.005 | $gm^{-2}$ |
| $r_N$ | 0.005 | $día^{-1}$ |
| $r_1$ | 0.3 | $día^{-1}$ |
| $k_1$ | 0.005 | $gm^{-2}$ |
| $k_2$ | 6 | $g^2m^{-4}$ |
| $k_3$ | 2 | $g^2m^{-4}$ |
| $f_1$ | 2 | $día^{-1}$ |
| $f_2$ | 10 | $día^{-1}$ |
| $\gamma$ | 0.02 | |
| $\eta$ | 0.5 | |
| $d_1$ | 0.1 | $día^{-1}$ |
| $d_2$ | 0.55 | $día^{-1}$ |
| $d_3$ | 0.5 | $día^{-1}$ |
| $d_4$ | 0.1 | $día^{-1}$ |
| $e_1$ | 0.001 | $día^{-1}$ |
| $e_2$ | 0.001 | $día^{-1}$ |
| $e_3$ | 0.001 | $día^{-1}$ |
| $e_4$ | 0.001 | $día^{-1}$ |

Tabla 3.1 Parámetros empleados en el modelo CASM

**Apéndice B**

Parámetros empleados para el modelo de oscilador glucolítico. Los valores fueron tomados de (Bier y colaboradores, 2000) y corresponden a un estado oscilante.

| Parámetros | Valores en unidades adimensionales |
|---|---|
| G(0) | 10.5 |
| T(0) | 0.04 |
| $V_{in}$ | 0.36 |
| $k_1$ | 0.02 |
| $k_p$ | 6 |
| $K_m$ | 13 |

Tabla 3.2 Parámetros del oscilador glucolítico.